\pdfoutput=1
\documentclass{article}
\usepackage[T1]{fontenc}
\usepackage{mathrsfs} 
\usepackage[utf8]{inputenc}
\usepackage{etex}
\makeindex
\newcommand{\BAF}{\mathbb{A}_{\fract,\FU}}
\usepackage{makecell}
\newcommand{\card}[1]{|#1|}
\newcommand{\bigO}[1]{O\left(#1\right)}
\usepackage{amsmath}
\usepackage{amsthm}
\usepackage{amssymb} 
\usepackage{tikz}
\usetikzlibrary{positioning,shadows,arrows,automata}
\DeclareMathOperator{\true}{True}
\DeclareMathOperator{\false}{False}
\newcommand{\BAR}{\mathbb{A}_{\R^{\ge0,\FU}}}
\newcommand{\del}[2]{\delta(#1,#2)}
\newcommand{\dell}[2]{\delta'(#1,#2)}
\usepackage{makeidx}

\newcommand{\realDot}{\star}
\newcommand{\wordToReal}[2][b]{\wordToNumber[#1]{#2}^{\R}}
\newcommand{\wordToFractional}[2][b]{\wordToNumber[#1]{#2}^{\fract}}
\newcommand{\wordToNatural}[2][b]{\wordToNumber[#1]{#2}^{\nat}}
\newcommand{\wordToNumber}[2][b]{\left[#2\right]_{#1}}
\newcommand{\setRealToLanguage}[2][b]{L_{#1}(#2)}
\newcommand{\set}[1]{\left\{ #1\right\}}
\newcommand{\ARPar}[1]{\mathcal{A}_{#1}}
\newcommand{\AR}{\ARPar{R}}
\newcommand{\FU}{\mathcal{S}}
\newcommand{\toInfWord}[1]{L_{\omega}\left({#1}\right)}
\newcommand{\nat}{I}
\newcommand{\fract}{F}
\newcommand{\natPart}[1]{{#1}_{\nat}}
\newcommand{\fraPart}[1]{{#1}_{\fract}}
\newcommand{\autPar}[5]{\left(#1,#2,#3,#4,#5\right)}
\newcommand{\aut}{\autPar Q\alphabet\delta{q_{0}}F}
\newcommand{\digitSet}[1][b]{\Sigma_{#1}}
\newcommand{\digitDotSet}[1][b]{\Sigma^{\realDot}_{#1}}
\newcommand{\alphabet}{A}
\newcommand{\emptyStates}[1][\mathcal{A}]{Q_{\emptyset,#1}}
\newcommand{\prefix}[2]{#1\left[<#2\right]}
\newcommand{\suffix}[2]{#1\left[\ge{}#2\right]}
\newcommand{\length}[1]{\left|{#1}\right|}
\newcommand{\inftyStates}[1][\mathcal{A}]{Q_{\infty,#1}}
\newcommand{\zuStates}[1][\mathcal{A}]{Q_{[0,1],#1}}
\newcommand{\natStates}[1][\mathcal{A}]{Q_{\nat,#1}}
\newcommand{\fraStates}[1][\mathcal{A}]{Q_{\fract,#1}}

\newcommand{\floor}[1]{\left\lfloor{#1}\right\rfloor}
\newcommand{\emptyState}[1][\mathcal{A}]{q_{\emptyset,#1}}
\newcommand{\inftyState}[1][\mathcal{A}]{q_{\infty,#1}}
\newcommand{\zuState}[1][\mathcal{A}]{q_{[0,1],#1}}
\newcommand{\iniState}{q_{0}}
\newcommand{\exRR}{\left(\frac13,2\right]\cup\left(\frac83,3\right]\cup\left(\frac{11}3,\infty\right]}
\newcommand{\exRfrac}{\left[\frac{1}{4},\frac{1}{3}\right)\cup\set{\frac{11}{24},\frac23}}
\newcommand{\suc}[1]{s_{#1}}

\usepackage{complexity}
\newcommand{\fo}[1]{\FO\left[#1\right]}
\newcommand{\ef}[1]{\sigF{1}{#1}}
\newcommand{\qf}[1]{\sigF{0}{#1}}
\newcommand{\sigF}[2]{\Sigma_{#1}\left[#2\right]}

\usepackage{hyperref}

\newtheorem{theorem}{Theorem}[section]
\newtheorem{lemma}[theorem]{Lemma}  
\newtheorem{proposition}[theorem]{Proposition}  
\newtheorem{corollary}[theorem]{Corollary}  
\theoremstyle{definition}
\newtheorem{example}[theorem]{Example}  
\newtheorem{definition}[theorem]{Definition}  
\begin{document}
\title{Büchi automata recognizing sets of reals definable in
  first-order logic with addition and order}
%
\author{Arthur Milchior}
%
%
\maketitle
\begin{abstract}
  This work considers weak deterministic Büchi automata reading  encodings of non-negative reals in a fixed base. A Real Number  Automaton is an automaton which recognizes all encoding of elements  of a set of reals.  It is explained how to decide in linear time{}  whether a set of reals recognized by a given minimal weak  deterministic RNA is $\textsc{FO}[\mathbb R;+,<,1]$-definable{}. Furthermore, it  is explained how to compute in quasi-quadratic (respectively,  quasi-linear) time an existential (respectively,  existential-universal) $\textsc{FO}[\mathbb R;+,<,1]$-formula which defines the  set of reals recognized by the automaton.  It is also shown that  techniques given by Muchnik and by Honkala for automata over vector  of natural numbers also works on vector of real numbers. It implies  that some problems such as deciding whether a set of tuples of reals  $R\subseteq\mathbb R^{d}$ is a subsemigroup of $(\mathbb R^{d},+)$ or is  $\textsc{FO}[\mathbb R;+,<,1]$-definable is decidable.
 \end{abstract}
%
%
%

\section*{Introduction}

This paper deals with logically defined sets of numbers encoded by
weak deterministic Büchi automata.  The sets of tuples of integers
whose encodings in base $b$ are recognized by a finite automaton are
called the $b$-recognizable sets.  By \cite{bruyere}, the
$b$-recognizable sets of vectors of integers are exactly the sets which are
$\fo{\mathbb Z;+,<,V_{b}}$-definable, where $V_{b}(n)$ is the greatest
power of $b$ dividing $n$. It was proven in
\cite{semenov-theorem,Cobham} that the $\fo{\mathbb N;+}$-definable
sets are exactly the sets which are $b$- and $b'$-recognizable for
every $b\ge 2$.

The preceding results naturally led to the following problem: deciding
whether a finite automaton recognizes a $\fo{\mathbb N;+}$-definable set of
$d$-tuples of integers for some dimension $d\in\mathbb N^{>0}$. In the case
of dimension $d=1$, the decidability was proven in \cite{Honkala86}.
For $d>1$, the decidability was proven in \cite{muchnik}. Another
algorithm was given in \cite{Leroux}, which solves this problem in
polynomial time. For $d=1$, a quasi linear time algorithm was given in
\cite{Sakarovitch}.

\paragraph{}
The above-mentioned results about sets of tuples of natural numbers
and finite automata have then been extended to results about set of
tuples of reals recognized by a Büchi automata.  The notion of Büchi
automata is a formalism which describes languages of infinite words,
also called $\omega$-words. The Büchi automata are similar to the
finite automata. The main difference between the two kinds of automata
is that finite automata accept finite words which admits runs ending
on accepting state, while Büchi automata accepts infinite words which
admit runs in which an accepting state appears infinitely often.

One of the main differences between finite automata and Büchi automata
is that finite automata can be determinized while deterministic Büchi
automata are less expressive than Büchi automata. For example, the
language $L_{\mbox{fin }a}$ of words containing a finite number of
times the letter $a$ is recognized by a Büchi automaton, but is not
recognized by any deterministic Büchi automaton. This statement
implies, for example, that no deterministic Büchi automaton recognizes
the set of reals of the form $nb^{p}$ with $n\in\mathbb N$ and $p\in\mathbb Z$, that
is, the reals which admits no encoding in base $b$ with a finite
number of non-0 digits.

Another main difference between the two classes of automata is that
the class of languages recognized by finite automata is closed under
complement while the class of languages recognized by deterministic
Büchi automata is not closed under complement. For example,
$L_{\mbox{inf }a}$, the complement of $L_{\mbox{fin }a}$, is recognized
by a deterministic Büchi automaton.

A Real Vector Automaton (RVA, See e.g. \cite{weak-R-+-vector}) of
dimension $d$ is a Büchi automaton $\mathcal{A}$ of alphabet
$\set{0,\dots,b-1}^{d}\cup\set\realDot$, which recognizes the set of
encoding in base $b$ of the elements of a set of vectors of reals.
Equivalently, for $w$ an infinite word encoding a vector of dimension
$d$ of real $\left(r_{0},\dots,r_{d-1}\right)$, if $w$ is recognized
by $\mathcal{A}$, then all encodings $w'$ of
$\left(r_{0},\dots,r_{d-1}\right)$ are recognized by $\mathcal{A}$. In
the case where the dimension $d$ is 1, those automata are called Real
Number Automata (RNA, See e.g. \cite{real}).

The sets of tuples of reals whose encoding in base $b$ is recognized
by a RVA are called the $b$-recognizable sets. By \cite{LinArCon},
they are exactly the $\fo{\mathbb R,\mathbb Z;+,<,X_{b},1}$-definable sets. The
logic $\fo{\mathbb R,\mathbb Z;+,<,X_{b},1}$ is the first-order logic over reals
with a unary predicate which holds over integers, addition, order, the
constant one, and the function $X_{b}(x,u,k)$. The function
$X_{b}(x,u,k)$ holds if and only if $u$ is equal to some $b^{n}$ with
$n\in\mathbb Z$ and there exists an encoding in base $b$ of $x$ whose digit in
position $n$ is $k$. That is, $u$ and $x$ are of the form:
\begin{eqnarray*}
  \begin{array}{lllllllllll}
    u=&0&\dots&0&\realDot&0&\dots&0&1&0&\dots\\
    x=& &\dots& &\realDot& &\dots& &k& &\dots
  \end{array}
\end{eqnarray*}
or of the form:
\begin{eqnarray*}
  \begin{array}{lllllllllll}
    u=&0&\dots&0&1&0&\dots&0&\realDot&0&\dots \\
    x=& &\dots& &k& &\dots& &\realDot& &\dots
  \end{array}
\end{eqnarray*}
A weak deterministic Büchi automaton is a deterministic Büchi
automaton whose set of accepting states is a union of strongly
connected components. A set is said to be weakly $b$-recognizable if
it is recognized by a weak automaton in base $b$.  By
\cite{weak-R-+-vector}, a set is $\fo{\mathbb R,\mathbb Z;+,<}$-definable if and
only if its set of encodings is weakly $b$-recognizable for all
$b\ge2$. The class of weak deterministic Büchi automata is less
expressive than the class of deterministic Büchi automata. For
example, the language $L_{\mbox{inf }a}$ of words containing an
infinite number of $a$ is recognized by a deterministic Büchi
automaton but is not recognized by any weak deterministic Büchi
automaton. This implies that, for example, no weak deterministic Büchi
automaton recognizes the set of reals which are not of the form
$nb^{p}$ with $n\in\mathbb N$ and $p\in\mathbb Z$, since those reals are the ones
whose encoding in base $b$ contains an infinite number of non-$0$
digits.  Furthermore, by \cite{minimal-buchi}, weak deterministic
Büchi automata can be efficiently minimized.
\paragraph{}

We now recall some results about the above-mentioned logic{}.
By \cite{elimFOr}, the logic $\fo{\mathbb R;+,<,1}$ admits quantifier
elimination. By \cite[Section 6]{elimination}, the set of reals which
are $\fo{\mathbb R;+,<,1}$-definable are the finite union of intervals with
rational bounds. Those sets are called the \emph{simple sets}.
\subsubsection*{Main results}
It is shown that ideas given in \cite{muchnik} and \cite{Honkala86} to
create algorithms to decide properties of automata over integers can
be adapted to decide properties of RVA. For examples, those ideas are
used in Section \ref{sec:method} to give algorithms which decide whether a
Büchi automaton recognizes a $\fo{\mathbb R;+,<,1}$-definable set of tuple of
reals, a $\fo{\mathbb R,\mathbb Z;+,<}$-definable set of tuple of real or a
subsemigroup of $(\mathbb R^{d},+)$ for some $d\in\mathbb N^{>0}$. However, those
algorithms are inefficient.

It is then shown in Section \ref{sec:real-aut} that it is decidable in
linear time whether a RNA recognizes a $\fo{\mathbb R;+,<,1}$-definable{} set, that is,
a{} simple set.  This algorithm does not return
any false positive on weak deterministic Büchi automata which are not
RNA. A false negative is also exhibited and it is explained why this
case is more complicated than the case of RNA.  A characterization of
the minimal weak RNA which recognizes simple sets is also given.

Note that, if an automaton recognizes a{} simple set $R$, that is{} a
finite union of intervals, the minimal number of intervals in the
union is not polynomially bounded by the number of states of the
automaton (this is shown in Example \ref{ex:A2-01}).  It is shown in
Section \ref{sec:aut->set} that an existential (respectively,
existential-universal) $\fo{\mathbb R;+,<,1}$-formula which defines $R$ is
computable in quasi-quadratic (respectively quasi-linear) time.
\section{Definitions}\label{sec:def}
The definitions used in this paper are given in this section.  Some
basic lemmas are also given. Most definitions are standard.
\subsection{Basic Notations}\label{sec:bas}
Let $\mathbb N$\index{N@$\mathbb N$}, $\mathbb
Z$\index{Z@$\mathbb
  Z$},
$\mathbb Q$ \index{Q@$\mathbb Q$} and $\mathbb R$ \index{R} denote the
set of non-negative integers, integers, rationals and reals,
respectively. For $R\subseteq \mathbb R$, let $R^{\ge 0}$ and $R^{>0}$
denote the set of non-negative and of positive elements of $R$,
respectively. Let $\omega$ be the cardinality of $\mathbb N$.  For
$n\in\mathbb N$, \index{n@$\left[n\right]$ for $n\in\mathbb N$.} let
$[n]$ represent $\set{0,\dots,n}$. For $a,b\in\mathbb R$ with
$a\le b$, let $[a,b]$ denote the closed interval
$\set{r\in\mathbb R\mid a\le r\le b}$, and let $(a,b)$ denote the open
interval $\set{r\in\mathbb R\mid a<r<b}$. Similarly, let $(a,b]$
(respectively, $[a,b)$) be the half-open interval equals to the union
of $(a,b)$ and of $\set b$ (respectively, $\set a$).  For
$r\in\mathbb R$ let \index{r@$\floor r$}$\floor r$ be the greatest
integer less than or equal to $r$.
\subsection{Finite and infinite words}
An alphabet is a finite set, its elements are called letters. A finite
(respectively infinite) word of alphabet $\alphabet$ is a finite
(respectively infinite) sequence of letters of $\alphabet$. That is, a
function from $[n]$ to $\alphabet$ for some $n\in\mathbb N$ (respectively from
$\mathbb N$ to $\alphabet$). A set of finite (respectively infinite) word of
alphabet $\alphabet$ is called a language (respectively, an
$\omega$-language) of alphabet $\alphabet$. The empty word is denoted
$\epsilon$.

Let $w$ be a word. Let $\length{w}\in\mathbb N\cup\set\omega$ denote the
length of $w$. For $v$ a finite word, let $u=vw$ be the concatenation
of $v$ and of $w$, that is, the word of length $\length{v}+\length{w}$
such that $u[i]=v[i]$ for $i<\length{v}$ and $u[\length{v}+i]=w[i]$
for $i<\length{w}$. For $n< \length{w}$, let $w[n]$ denote the $n$-th
letter of $w$. Let $\prefix w{n}$ denote the \emph{prefix} of $w$ of
length $n$, that is, the word $u$ of length $n$ such that $w[i]=u[i]$
for all $i\in[n-1]$. Similarly, let $\suffix w{n}$ denote the suffix
of $w$ without its $n$-th first letters, that is, the word $u$ such
that $u[i]=w[i+n]$ for all $i\in[n-n]$. Note that
$w=\prefix wi\suffix wi$ for all $i<\length{w}$.

Let $L$ be a language of finite word and let $L'$ be either an
$\omega$-languages or a language of finite words. Let $LL'$ be the set
of concatenations of the words of $L$ and of $L$. For $i\in\mathbb N$, let
$L^{i}$ be the concatenations of $i$ words of $L$.  Let
$L^{*}=\bigcup_{i\in\mathbb N} L^{i}$ and
$L^{+}=\bigcup_{i\in\mathbb N^{>0}} L^{i}$. If $L$ is a language which does
not contains the empty word, let $L^{\omega}$ be the set of infinite
sequences of elements of $L$.

\subsubsection{Encoding of real numbers}
Let us now consider the encoding of numbers in an integer base
$b\ge2$. Let $\digitSet$ be equal to $[b-1]$, it is the set of digits
and let $\digitDotSet=\digitSet\cup\set\realDot$. The base $b$ is fixed
for the remaining of this paper. Two alphabets are considered in this
paper: $\digitSet$ and $\digitDotSet$. 

\index{.@$\wordToNumber{.}$}\label{def:wordToNumber}\index{Natural
  part of a word of $\digitSet^{*}\realDot\digitSet^{\omega}$}
\index{Fractional part of a word of
  $\digitSet^{*}\realDot\digitSet^{\omega}$} Let $\wordToNumber{.}$
denote the function which sends a finite or infinite word of alphabet
$\digitDotSet$ to the integer or to the real it represents. Formally,
for $w\in\digitSet^{*}$:
  \begin{equation*}
    \wordToNatural{w}=\sum_{i=0}^{\length{w}-1}b^{\length{w}-1-i}w[i].
  \end{equation*}
  For $w\in\digitSet^{\omega}$, 
  \begin{equation*}
    \wordToReal{w}=\sum_{i\in\mathbb N}b^{-i-1}w[i].
  \end{equation*}
  Let $w$ be an $\omega$-word with exactly one $\realDot$. It is of
  the form $w=\natPart{w}\realDot \fraPart{w}$, with
  $\natPart{w}\in\digitSet^{*}$ and
  $\fraPart{w}\in\digitSet^{\omega}$. The word $\natPart{w}$ is called the
  natural part of $w$ and the $\omega$-word $\fraPart{w}$ is called its
  fractional part. Then :
  \begin{equation*}
    \wordToReal{\natPart{w}\realDot \fraPart{w}}=\wordToNatural{\natPart{w}}+\wordToFractional{\fraPart{w}}.
  \end{equation*}
  Finally, $\wordToReal{w}$ is undefined if $w$ contains at least two
  letters $\realDot$.
  There is no ambiguity in the definition of $\wordToNatural{\cdot}$
  since the four domains of definitions partition
  $\left(\digitDotSet\right)^{\omega}$.  Note that
  $\wordToReal{\natPart{w}}\in\mathbb N$, $\wordToReal{\fraPart{w}}\in[0,1]$ and
  $\wordToFractional w=\wordToReal {\natPart{w}}+\wordToReal{\fraPart{w}}$.
  Examples of numbers with their base $2$ encodings are now given.
\begin{example}
  \begin{equation*}
    \arraycolsep=0.5pt
    \begin{array}{rclp{2mm}rclp{2mm}rclp{2mm}rclp{2mm}rcl}
      \wordToReal[2]{(10)^{\omega}}&=&\frac 23 &
      &\wordToReal[2]{(01)^{\omega}}&=&\frac13&
      &\wordToReal[2]{0(10)^{\omega}}&=&\frac13&
      &\wordToReal[2]{0(1)^{\omega}}&=&\frac{1}{2}&
      &\wordToReal[2]{1(0)^{\omega}}&=&\frac12
      \\
      \wordToReal[2]{10}&=&2&
      &\wordToReal[2]{1}&=&1&
      &\wordToReal[2]{01}&=&1&
      &\wordToReal[2]{\epsilon}&=&0&
      &\wordToReal[2]{00000}&=&0
      \\
      \multicolumn{5}{r}{\wordToReal[2]{10\realDot(10)^{\omega}}}&=&\frac{8}{3}&
      &\wordToReal[2]{\realDot0(1)^{\omega}}&=&\frac{1}{2}&
      &\multicolumn{5}{r}{\wordToReal[2]{00000\realDot 1(0)^{\omega}}}&=&\frac{1}{2}.
    \end{array}
  \end{equation*}
\end{example}
Some properties of concatenation and encodings of reals are now
stated. The proof of the lemma is straightforward from the definition.
\begin{lemma}
  Note that for all $v\in\digitSet^{*}$, $w\in\digitSet^{\omega}$ and
  $a\in\digitSet$:
  \begin{equation*}
    \begin{array}{rclcrclp{2mm}rcl}
      \wordToReal{aw}&=&\frac{a+\wordToReal{w}}{b},&
      &\wordToNatural{av}&=&ab^{\length v}+\wordToNatural{v},&
      &\wordToReal{w}&=&\wordToReal{0\realDot w},\\
      \wordToNatural{va}&=&b\wordToNatural{v}+a,
      &\mbox{ and }
               &\wordToReal{av\realDot w}&=&ab^{\length v}+\wordToReal{v\realDot w}.
    \end{array}
  \end{equation*}
\end{lemma}

\subsubsection{Encoding of rationals}\label{sec:rati}
In this section, some basic facts about rationals are recalled (see
e.g. \cite{number-theory-hardy}).  The rationals are exactly the
numbers which admit encodings in base $b$ of the form
$u\realDot{}vw^{\omega}$ with $u,v\in\digitSet^{*}$ and
$w\in\digitSet^{+}$.  Rationals of the form $nb^{p}$, with $n\in\mathbb N$
and $p\in\mathbb Z$, admit exactly two encodings in base $b$ without
leading 0 in the natural part.  If $p<0$, the two encodings are
of the form $u\realDot{}va(b-1)^{\omega}$ and
$u\realDot{}v(a+1)0^{\omega}$, with $u,v\in\digitSet^{*}$ and
$a\in[b-2]$. Otherwise, if $p\ge 0$, the two encodings are of
the form $ua(b-1)^{q}\realDot{}(b-1)^{\omega}$ and
$u(a+1)0^{q}\realDot{}0^{\omega}$ with $u\in\digitSet^{*}$,
$a\in[b-2]$ and $q\in\mathbb N$.  The rationals which are not of the form
$n b^{p}$ admit exactly one encoding in base $b$ without leading 0
in the natural part.
\subsubsection{Encoding of sets of reals}\label{subsec:rep-set-real}
In this section, relations between languages and set of reals are
recalled.

Given a language $L$ in $\digitSet^{\omega}$ or in
$\digitSet^{*}\realDot\digitSet^{\omega}$, let
\index{$\wordToReal{L}$}$\wordToReal{L}$ be the set of reals admitting
an encoding in base $b$ in $L$. The language $L$ is said to be an encoding in base $b$ of the set of reals $\wordToReal L$.
Reciprocally, given a set $R\subseteq\mathbb R^{\ge0}$ of reals,
$\setRealToLanguage R$ is the set of all encodings in base $b$ of the
elements of $R$. 

Following \cite{Leroux}, a language $L$ is said to be
\emph{saturated}\index{Saturated} if for any number $r$ which admits an encoding in base $b$ in $L$, all encoding in base $b$ of $r$ belongs
to $L$. The saturated languages are of the form $\setRealToLanguage R$
for $R\subseteq\mathbb R^{\ge0}$.  Note that
$\wordToReal{\setRealToLanguage{R}}=R$ for all sets
$R\subseteq\mathbb R^{\ge0}$. Note also that
$L\subseteq\setRealToLanguage{\wordToReal{L}}$, and the subset
relation is an equality if and only if $L$ is saturated. In general, a
set of reals may have infinitely many encodings in base $b$. For
example, for $I\subseteq\mathbb N$ an arbitrary set,
$\set{0,1}^{\omega}\setminus\set{0^{i}1^{\omega}\mid i\in I}$ is an encoding in base 2 of the language of the simple set $[0,1]$.  An
example of set of reals is now given.
\begin{example}
  Let $L=\digitSet[2]^{*}0^{\omega}$ and
  $L'=\digitSet[2]^{*}0\digitSet[2]^{*}(0^{\omega}+1^{\omega})$. Both
  $\wordToReal[2]{L}$ and
  $\wordToReal[2]{L'}$ are
  $R=\set{\frac{n}{2^{p}}\mid{}n,p\in\mathbb N,n<2^{p}}$, but only $L'$ is
  saturated. Therefore $\setRealToLanguage[2]{R}=L'$.
\end{example}

\subsection{Deterministic Büchi automata}
This paper deals with {Deterministic }Büchi automata. This
notion is now defined.

A \emph{{Deterministic }Büchi automaton}\index{{Deterministic }Büchi
  automaton} is a 5-tuple
$ \mathcal{A}=\autPar{Q}{\alphabet}{\delta}{q_0}{F} $, where $Q$ is a
finite \emph{set of states}, $\alphabet$ is an alphabet,
$\delta\subseteq Q\times\alphabet\times Q$ is the \emph{transition
  relation}, ${q_0\in Q}$ is the {\emph{initial states}} and
$F\subseteq Q$ is the set of \emph{accepting states}. A state
belonging to $Q\setminus F$ is said to be a \emph{rejecting
  state}.\index{Rejecting state}


An example of deterministic Büchi automaton is now given. This example
is used thorough this paper to illustrate properties of Büchi
automaton reading set of real numbers.
\begin{example}\label{ex:unbounded}
  Let $R=\exRR$. The set of encodings in base 2 of reals of $R$ is
  recognized by the automaton pictured in Figure \ref{fig:ex-AR-R>0}.
   \begin{figure}[!h]
    \centering
      \begin{tikzpicture}[->, >=stealth', shorten >=1pt, auto, node
        distance=1.8cm, thick, main node/.style={circle, fill=!20, draw,
          font=\sffamily\Large\bfseries, inner sep=0.1pt, minimum
          size=1cm}] \tikzset{every state/.style={minimum size=1cm}}
        
        \node[state, initial, initial text={}]        (0)       {$\iniState$};
        \node[state, right of=0]       (1)     {$q_{1}$};
        \node[state, right of=1]       (2)     {$q_{2}$};
        \node[state, right=4.2cm of 2] (3)     {$q_{3}$};
        \node[state, right=2.6cm of 3] (infty) {$\infty$};
        \node[state, below of=3]            (3epsilon) {$(3,\epsilon)$};
        \node[state, below of=3epsilon]     (31)       {$(3,1)$};
        \node[state, right of=31]           (310)      {$(3,10)$};
        \node[state, right of=3epsilon,accepting] (30)       {$(3,0)$};
        \node[state, below of=2]            (2epsilon) {$(2,\epsilon)$};
        \node[state, below of=2epsilon]     (21)       {$(2,1)$};
        \node[state, right of=21]           (210)      {$(2,10)$};
        \node[state, right of=2epsilon,accepting] (20)       {$(2,0)$};
        \node[state, below of=0]     (0epsilon) {$(0,\epsilon)$};
        \node[state, below of=0epsilon]     (00) {$(0,0)$};
        \node[state, right of=210,accepting] (f) {$\zuState$};
        
        \path[every node/.style={font=\sffamily\small}] 
        (0)    edge [loop above] node  {0} (0)
               edge node  {1} (1)
               edge node {$\realDot$} (0epsilon)
        (1)    edge node  {0} (2)
               edge [bend left] node  {1} (3)
        (infty)edge [loop right]  node {0,1} (infty)
        (2)    edge [bend left]node {0,1} (infty)
               edge node {$\realDot$} (2epsilon)
        (3)    edge node {0,1} (infty)
               edge node {$\realDot$} (3epsilon)
        (0epsilon) edge [bend right] node {0} (00)
        (00) edge [bend right]node {1} (0epsilon)
        (2epsilon) edge node {0} (20)
                   edge node {1} (21) 
        (20) edge [loop above] node {0} (20)
        (21) edge [bend left]node {0} (210)
        (210) edge [bend left]node {1} (21)
        (3epsilon) edge node {1} (31)
                   edge node {0} (30) 
        (30) edge [loop above] node {0} (30)
        (31) edge [bend left]node {0} (310)
              edge node {1} (f)
        (310) edge [bend left]node {1} (31)
        (f) edge [loop above] node {0,1} (f)
        ;
        \draw (infty) -- node [pos=.2] {$\realDot$} (11.5,-4.8) --  (7.8,-4.8) -- (f) ;
        \draw[-] (1) -- node [pos=.2] {$\realDot$} +(0,-4.8) ;
        \path[-] (21) edge node [pos=.5,left] {1} (4,-4.8) ;
        \draw (0epsilon) -- node [left] {1}  (-1,-3) -- (-1,-4.8) -- (7,-4.8) -- (f);

      \end{tikzpicture}
      \caption{Automaton $\AR$ of Example \ref{ex:unbounded}}
      \label{fig:ex-AR-R>0}
    \end{figure}
\end{example}

{From now on in this paper, all automata are assumed to be
  deterministic. The function $\delta$ is implicitly extended on
  $Q\times \alphabet^{*}$ by $\del{q}{\epsilon}=q$ and
  $\del{q}{aw}=\del{\del{q}{a}}{w}$ for $a\in\alphabet$ and
  $w\in\alphabet^{*}$.  }

\index{Run of an automaton} Let $\mathcal{A}$ be an automaton and $w$ be an
infinite word. A \emph{run}\index{Run} $\pi$ of $\mathcal{A}$ on $w$ is a
mapping $\pi:\mathbb N\mapsto Q$ such that $\pi(0)=\iniState$ and
$\del{\pi(i)}{w[i]}={\pi(i+1)}$ for all $i<\length{w}$. Let $inf(\pi)$
be the set of states of $Q$ that occur infinitely often in the run
$\pi$. A run $\pi$ on an $\omega$-word is said to be accepting if
$inf(\pi)\cap F\ne\emptyset$. Equivalently, the run is accepting if
there exists a state $q\in F$ such that there is an infinite number of
$i\in\mathbb N$ such that $\pi(i)=q$.
Example \ref{ex:unbounded} is now resumed.
\begin{example}
  Let $\mathcal{A}$ be the automaton pictured in Figure \ref{fig:ex-AR-R>0}.  The run of $\mathcal{A}$ on
  $011\realDot(10)^{\omega}$ is \begin{equation*}
    \left(q_{0},q_{0},q_{1},q_{3},(3,\epsilon),(3,1),(3,10),\dots\right)
  \end{equation*}
  with the two last states repeated infinitely often.  The Büchi
  automaton $\mathcal{A}$ does not accept $011\realDot(10)^{\omega}$ since
  this run does not contain any accepting state.

  The run of $\mathcal{A}$ on $\realDot1^{\omega}$ is $
  \left(q_{0},(0,\epsilon),\zuState,\dots\right)$
  with the last state repeated infinitely often. The Büchi automaton
  $\mathcal{A}$ accepts $\realDot1^{\omega}$ since the accepting state
  $\zuState$ appears infinitely often in the run.
\end{example}

Let $\mathcal{A}$ be a finite automaton. Let
\index{A@$\toInfWord{\mathcal{A}}$}$\toInfWord{\mathcal{A}}$ be the set of infinite
words $w$ such that a run of $\mathcal{A}$ on $w$ is accepting.  An
$\omega$-language is said to be \emph{recognizable}\index{Recognizable
  language} if it is recognized by a Büchi automaton.
Example \ref{ex:unbounded} is now resumed.
\begin{example}
  Let $\mathcal{A}$ be the Büchi automaton pictured in Figure \ref{fig:ex-AR-R>0}.  It recognizes the language of encodings in
  base 2 of the reals of $\exRR$. It is explained in Example
  \ref{ex:construct-RR} how this automaton was computed.
\end{example}
For $q\in Q$, let $\mathcal{A}_{q}$ \index{A@$\mathcal{A}_{q}$ for
  $\mathcal{A}$ an automaton and $q$ a state.} be
$\left(Q_{q},\alphabet,\delta,q,F_{q}\right)$, where
$Q_{q}$\index{Q@$Q_{q}$ for $Q$ a set of states and $q$ a state.}  is
the set of states of $Q$ accessible from $q$, and $F_{q}=F\cap Q_{q}$.

Note that, if there are no finite word $w$ such that
$\del{\iniState}{w}=\iniState$, then $Q_{q}\subsetneq Q$ for all
$q\ne \iniState$. Note also that, if $w\in\alphabet^{*}$ is such that
$\del{q_{0}}{w}=q$ then a word $w'\in\alphabet^{\omega}$ is accepted
by $\mathcal{A}_{q}$ if and only if $ww'$ is accepted by $\mathcal{A}$.

\subsubsection{Accessibility and recurrent states}
When the notions of initial and of accepting states are ignored, an
automaton can be considered as a directed labelled graph. Some
definitions related to this  graph are introduced in this section.

A state $q$ is said to be \emph{accessible}\index{Accessible} from a
state $q'$ if there exists a finite non-empty word $w$ such that
$\del{q'}{w}={q}$. Following \cite{minimal-buchi}, a state $q$ is said
to be \emph{recurrent}\index{Recurrent state} if it is accessible from
itself and \emph{transient} otherwise\index{Transient
  state}. Transient states are called \emph{trivial} in
\cite{Boigelot2007}. The \index{Strongly connected
  components}\emph{strongly connected component} of a recurrent state
$q$ is the set of states $q'$ such that $q'$ is accessible from $q$
and $q$ is accessible from $q'$. A strongly connected component $C$ is
said to be a \index{Leaf}leaf if for all $a\in\alphabet$, for all
$q\in C$, $\del{q}{a}\in C$.  Let $C$ be a strongly connected
component. It is said to be a cycle if for each $q\in C$, there exists
a unique $\suc{q}\in\alphabet$ such that $\del{q}{\suc{q}}\in C$.
Example \ref{ex:unbounded} is now resumed.
\begin{example}\label{ex:AR-R>0-min}
  The transient states of the automaton pictured in Figure
  \ref{fig:ex-AR-R>0-min} are $q_{0}$, $q_{1}$, $q_{2}$, $q_{3}$,
  $(2,\epsilon)$ and $(3,\epsilon)$. All other states are
  recurrent. The cycles are $\set{q_{0}}$, $\set{(0,\epsilon),(0,0)}$,
  $\set{(2,0)}$, $\set{(2,1), (2,10)}$, $\set{(3,0)}$ and
  $\set{(3,1), (3,10)}$. The strongly connected component which are
  not cycles are $\emptyState$, $\inftyState$ and $\zuState$.
\end{example}

The following lemma allows to consider recurrent states in any run which
is long-enough.
\begin{lemma}\label{lem:run-recurrent}
  Let $\mathcal{A}$ be a Büchi automaton with $n$ states let $w$ be an
  $\omega$-word and let $\pi$ be the run of $\mathcal{A}$ on $w$. Let
  $N\subseteq \mathbb N$ be a set of cardinal at least $(n+1)$. Then there is
  $i<i'$ belonging to $N$ such that $\pi(i)=\pi(i')$ is a recurrent
  state.
\end{lemma}
\begin{proof}
  Since the cardinality of $N$ is greater than the number of state, by
  the pigeonhole principle, there exists $i<i'$ belonging to $N$ such
  that $\pi(i)=\pi(i')$.  Let $w'$ be the factor of $w$ containing the
  letters $i+1$ to $i'$, then $\del{\pi(i)}{w'}=\pi(i')=\pi(i)$,
  therefore, the state $\pi(i)$ is recurrent, with $i$ belonging to
  $N$.
\end{proof}
\subsubsection{Quotients, Morphisms and Weak Büchi Automata}
In this section, the notion of quotient of automata and of morphism of
automata are introduced. A class of automata admitting minimal
quotient is then introduced.

\begin{definition}[Morphism of Büchi automata,
  Quotient]\label{def:morph}\index{Morphism of
    automata}\index{Quotient of automata} Let $\mathcal{A}=\aut$ and
  $\mathcal{A}'=\autPar{Q'}{\alphabet}{\delta'}{\iniState'}{F'}$ be
  two Büchi automata over the same alphabet. A surjective function
  $\mu:Q\to Q'$ is a \emph{morphism}\index{Morphism of automata} of
  Büchi automata if and only if:
  \begin{enumerate}
  \item $\mu(\iniState)=q'_{0}$,
  \item for each $q\in Q$,
    $\toInfWord{\mathcal{A}_{q}}\ne\toInfWord{\mathcal{A}_{q'}}$.
  \end{enumerate}
  The Büchi automaton $\mathcal{A}'$ is said to be a \emph{quotient}
  of $\mathcal{A}$ if there exists a morphism from $\mathcal{A}$ to
  $\mathcal{A}'$.
\end{definition}
The notion of minimal Büchi automaton is now introduced.
\begin{definition}[Minimal Büchi automaton]\label{Minimal Büchi automaton}
  Let $\mathcal{A}=\autPar{Q}{\digitDotSet}{\delta}{\iniState}{F}$ be
  a Büchi automaton. It is said to be minimal if for each distinct
  states $q,q'\in Q$,
  $\toInfWord{\mathcal A_{q}}=\toInfWord{\mathcal A_{q'}}$.
\end{definition}
In general, Büchi automata does not admit minimal quotient. A class of
Büchi automata admitting minimal quotient is now introduced.
\begin{definition}[Weak automata]
  Let $\mathcal{A}=\autPar{Q}{\digitDotSet}{\delta}{\iniState}{F}$ be
  a Büchi automaton. It is said to be \emph{weak} if for each
  recurrent accepting state $q$ of $\mathcal{A}$, all states of the
  strongly connected components of $q$ are accepting.

  An $\omega$-language is said to be \emph{weakly
    recognizable}\index{Weakly recognizable language} if it is
  recognized by a weak Büchi automaton.
\end{definition}
 The main theorem concerning
quotient of weak Büchi automata is now recalled.
\begin{theorem}[\cite{minimal-buchi}]
  Let $\mathcal{A}$ be a weak Büchi automaton with $n$ states such
  that all states of $\mathcal{A}$ are accessible from its initial
  state. Let $c$ be the cardinality of $\alphabet$.  There exists a
  minimal weak Büchi automaton $\mathcal{A}'$ such that there exists a
  morphism of automaton $\mu$ from $\mathcal{A}$ to
  $\mathcal{A}'$. The automaton $\mathcal{A}'$ and the morphism $\mu$
  are computable in time $\bigO{n\log(n)c}$ and space $\bigO{nc}$.
\end{theorem}
It follows easily from Property \eqref{def:morph-trans} that, for all
$w\in\alphabet^{*}$, $\dell{\mu(q)}{w}=\mu(\del{q}{w})$.
Example \ref{ex:unbounded} is now resumed.
\begin{example}\label{ex:unbounded-min}
  Let $\AR$ be the automaton pictured in Figure \ref{fig:ex-AR-R>0}. Its
  minimal quotient is pictured in Figure \ref{fig:ex-AR-R>0-min}.  
  \begin{figure}
    \centering
    \begin{tikzpicture}[->, >=stealth', shorten >=1pt, auto, node
      distance=2cm, thick, main node/.style={circle, fill=!20, draw,
        font=\sffamily\Large\bfseries, inner sep=0.1pt, minimum
        size=30cm}] \tikzset{every state/.style={minimum size=1.2cm}}
      
      \node[state, initial, initial text={}]        (0)       {$\iniState$};
      \node[state, right of=0]     (2) {$q_{2}$};
      \node[state, right=3.7cm  of 2]     (infty) {$\inftyState[R]$};
      \node[state, right=1.3cm of infty]     (1) {$q_{1}$};
      \node[state, below of=2]     (2epsilon) {$(2,\epsilon)$};
      \node[state, accepting, right of=2epsilon] (20) {$(2,0)$};
      \node[state, below of=0]     (0epsilon) {$(0,\epsilon)$};
      \node[state, left of=0epsilon]     (00) {$(0,0)$};
      \node[state, below of=infty,accepting] (f) {$\zuStates[R]$};
      
      \path[every node/.style={font=\sffamily\small}] 
      (0)    edge [loop above] node  {0} (0)
             edge [bend left,in=140,out=40]node  {1} (1)
             edge node {$\realDot$} (0epsilon)
      (1)    edge [bend right] node [above] {0,1} (2)
             edge node {$\realDot$} (f)
      (infty)edge [loop right]  node {0,1} (infty)
             edge node {$\realDot$} (f)
      (2)    edge node {0,1} (infty)
             edge node {$\realDot$} (2epsilon)
      (0epsilon) edge[bend left] node {0} (00)
                 edge [bend right] node {1} (f)
      (00) edge [bend left]node {1} (0epsilon)
      (2epsilon) edge  node {1} (0epsilon)
                 edge  node {0} (20)
      (20) edge [loop above] node {0} (20)
      (f) edge [loop left] node {0,1} (f)
      ;
    \end{tikzpicture}
    \caption{Minimal quotient of automaton $\AR$ of
      Figure \ref{fig:ex-AR-R>0}}
    \label{fig:ex-AR-R>0-min}
  \end{figure}
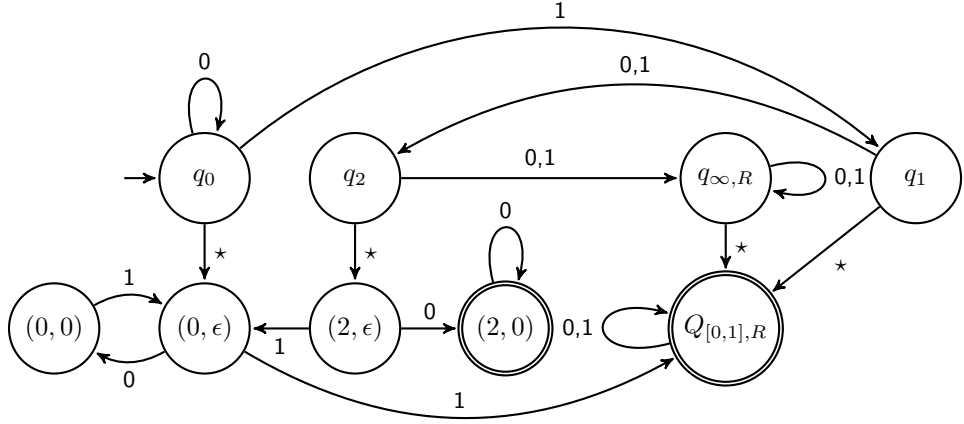
\end{example}

The following lemma shows that each strongly connected component of a
quotient by a morphism $\mu$ from an automaton $\mathcal{A}$ is the image of
a strongly connected component of $\mathcal{A}$.
\begin{lemma}\label{lem:scc-morphism-codomain}
  Let $\mathcal{A}=\aut$ and
  $\mathcal{A}'=\autPar{Q'}{\digitSet}{\delta'}{q'_{0}}{F'}$ be two Büchi
  automata. Let $\mu$ be a morphism from $\mathcal{A}$ to $\mathcal{A}'$. Let $C'$
  be a strongly connected component of $\mathcal{A}'$. There exists a
  strongly connected component $C\subseteq Q$ such that $\mu(C)=C'$
  and such that, for all $q\in Q\setminus C$ accessible from $C$,
  $\mu(q)\not\in C'$.
\end{lemma}
In order to prove this lemma, two other lemmas are required.
\begin{lemma}\label{lem:scc-morphism-codomain-inclusion}
  Let $\mathcal{A}$, $\mathcal{A}'$, $C'$ and $\mu$ as in
  Lemma \ref{lem:scc-morphism-codomain}.  Let $C$ be a strongly connected
  component of $\mathcal{A}$. Either $\mu(C)\cap C'=\emptyset$ or $\mu(C)\subseteq C'$.
\end{lemma}
\begin{proof}
  Let us assume that $\mu(C)\cap C'\ne\emptyset$ and let us prove that
  $\mu(C)\subseteq C'$. That is, let $q\in C$ and let us prove that
  $\mu(q)\in C'$. 
  
  Since $\mu(C)\cap C'\ne\emptyset$, there exists
  $q'\in\mu(C)\cap C'$.  Since $q'\in\mu(C)$, there exists $p\in C$
  such that $\mu(p)=q'$.  Since $p$ and $q$ belong to the same
  strongly connected component, there exists two non-empty finite
  words $v$ and $w$ such that $\del{p}{v}=q$ and
  $\del{q}{w}=p$. Therefore $\dell{\mu(p)}{v}=\mu(\del{p}{v})=\mu(q)$
  and $\dell{\mu(q)}{w}=\mu(\del{q}{w})=\mu(q)$. Therefore $\mu(q')$
  is accessible from $\mu(q)$ and $\mu(q')$ is accessible from
  $\mu(q)$. Hence $\mu(q)$ belongs to the strongly connected component
  of $p$. That is, $\mu(q)$ belongs to $C'$.
\end{proof}
\begin{lemma}\label{lem:scc-morphism-codomain-follow}
  Let $\mathcal{A}$, $\mathcal{A}'$, $C'$ and $\mu$ as in
  Lemma \ref{lem:scc-morphism-codomain}.  Let $q\in Q$ such that
  $\mu(q)\in C'$.  There exists a strongly connected component $C$ of
  $\mathcal{A}$, accessible from $q$, such that $\mu(C)\subseteq C'$.
\end{lemma}
\begin{proof}
  Since $\mu(q)\in C'$, the state $\mu(q)$ is recurrent, therefore
  there exists a non-empty word $w$ such that
  $\dell{\mu(q)}{w}=\mu(q)$. Let us prove by induction on $i\in\mathbb N$
  that $\mu(\del{q}{w^{i}})=\mu(q)$. The case $i=0$ is trivial, let us
  assume that the hypothesis holds for $i\in\mathbb N$ and let us prove that
  the induction hypothesis holds for $i+1$. It suffices to see that
  \begin{equation*}
    \mu(\del{q}{w^{i+1}})=
    \dell{\mu(q)}{w^{i+1}}=
    \dell{\dell{\mu(q)}{w^{i}}}{w}=
    \dell{\mu(q)}{w}=
    \mu(q)
  \end{equation*}
  By Lemma \ref{lem:run-recurrent}, there exists $i\in\mathbb N$ such that
  $\del{q}{w^{i}}$ is recurrent. Let $C$ be the strongly connected of
  $\del{q}{w^{i}}$. Since $\del{q}{w^{i}}\in C$ and
  $\mu(\del{q}{w^{i}})\in C'$, by
  Lemma \ref{lem:scc-morphism-codomain-inclusion}, it implies that
  $\mu(C)\subseteq C'$. Since $C$ is accessible from $q$ and
  $\mu(C)\subseteq C'$, the lemma is satisfied.
\end{proof}
Lemma \ref{lem:scc-morphism-codomain} is now proven.
\begin{proof}[Proof of Lemma \ref{lem:scc-morphism-codomain}]
  Let $\mu^{-1}(C')\subseteq Q$ be the set of states $q$ such that
  $\mu(q)\in C'$.  By definition of morphism, $\mu$ is surjective,
  hence $\mu^{-1}(C')$ is not empty. Let $q$ be a state belonging to
  $\mu^{-1}(C')$. By Lemma \ref{lem:scc-morphism-codomain-follow}, it
  implies that there exists a strongly connected component
  $C\subseteq\mu^{-1}(C')$.  Since $\mu^{-1}(C')$ is finite, there
  exists a strongly connected component $C$ such that no other
  strongly connected component of $\mu^{-1}(C')$ is accessible from
  $C$.  By Lemma \ref{lem:scc-morphism-codomain-follow} it implies that
  no state of $\mu^{-1}(C')\setminus C$ is accessible from $C$.

  Let us prove that $\mu(C)=C'$. Since, by hypothesis
  $\mu(C)\subseteq C'$, it remains to prove that $\mu(C)\supseteq C'$.
  Let $q'\in C'$ and let us prove that $q'\in\mu(C)$. Let $q\in C$.
  By hypothesis, $\mu(C)\subseteq C'$, therefore $\mu(q)\in C'$. Since
  $\mu(q)$ and $q'$ belong to the same strongly connected component
  $C'$, there exists a finite word $w$ such that
  $\dell{\mu(q)}{w}=q'$.  Then
  $\mu(\del{q}{w})=\dell{\mu(q)}{w}=q'$. Since
  $\mu(\del{q}{w})=q'\in C'$, $\del{q}{w}\in\mu^{-1}(C')$.  Since
  $\del{q}{w}\in \mu^{-1}(C')$ and $\del{q}{w}$ is accessible from
  $q\in C$, by hypothesis on $C$, $\del{q}{w}\in C$. Therefore
  $q'=\mu(\del{q}{w})\in\mu(C)$.
\end{proof}

\subsection{Logic}
The logic{} $\fo{\mathbb R;+,<,1}${} used
in this paper {is} introduced in this section. Note that, in
order to avoid ambiguity between the mathematical equality and the
formal equality of the logic, the symbol $\doteq$ is used in
first-order formulas.

Intuitively, $\FO$ stands for first-order. The first parameter $\mathbb R$
means that the (free or quantified) variables are interpreted by real
numbers. The $+$ and $<$ symbols mean that the function addition and
the binary order relation over reals can be used in formulas. Finally,
the last term, $1$, means that the only constant is 1.  The logic
$\fo{\mathbb R;+,<,1}$ is denoted by $\mathscr{L}$ in \cite{elimFOr}, where
it is proved that this logic admits quantifier elimination.  In this
paper, most results deal with the quantifier-free, the existential
fragment and the existential-universal fragment of $\fo{\mathbb R;+,<,1}$
denoted by $\qf{\mathbb R;+,<,1}$, $\ef{\mathbb R;+,<,1}$ and
$\sigF{2}{\mathbb R;+,<,1}$ respectively.

In the remaining of the paper, rationals are also used in the
formulas. It does not change the expressivity, as all rational
constants are $\qf{\mathbb R;+,1}$-definable.
Let $\phi\in\fo{\mathbb R;+,<,1}$. The length of $\phi$, denoted by
$\length{\phi}$\index{Length of formulas}, is recursively defined as
follows:
\begin{itemize}
\item The lengths of the constant \emph{$\frac{p}{q}$} is
  $\log{p+1}+\log{q}$.
\item The length of a sum $t_{1}+t_{2}$ is
  $1+\length{t_{1}}+\length{t_{2}}$.
\item The length of a multiplication by a rational constant
  $\frac pqt$ is $\length{\frac pq}+\length{t}$.
\item The length of an (in)equality is the sum of the length of the
  terms on both side, plus one, that is
  $\length{t_{1}<t_{2}}=\length{t_{1}\doteq
    t_{2}}=1+\length{t_{0}}+\length{t_{1}}$.
\item The length of Boolean combination and of quantification are 1
  plus the length of its subterms, that is
  $\length{\phi\lor\psi}=\length{\phi\land\psi}=1+\length{\phi}+\length{\psi}$
  and
  $\length{\exists x.\phi}=\length{\forall
    x.\phi}=\length{\neg\phi}=1+\length{\phi}$.
  \end{itemize}
\subsubsection{First-order definable sets of reals} In this section,
notations are introduced for the{} kind{} of sets
studied in this paper: the $\fo{\mathbb R;+,<,1}$-definable{} sets.

\label{sec:FU}Following \cite[Section 6]{elimination}, the
$\fo{\mathbb R;+,<,1}$-definable sets are called the \index{Simple
  sets}\emph{simple sets}. By \cite[Section 6]{elimination}, those
sets are the finite union of intervals with rational bounds. It
implies that there exists an integer $t_{R}$ such that for all
$x,y\ge t_{R}$, $x$ belongs to $R$ if and only if $y$ belongs to $R$.
\index{Threshold of a simple set.} The least such integer $t_{R}$ is
called the \emph{threshold of $R$}.

Note that every closed and half-closed intervals is the union of an
open interval and of singletons, hence it can be assumed that any
simple set $R$ is of the form
\begin{equation*}
  R=\bigcup_{i=0}^{I-1}(\rho_{i,\mathfrak L},\rho_{i,\mathfrak R})\cup\bigcup_{i=0}^{J-1}\set{\rho_{i,\mathfrak S}},
\end{equation*}
with $\rho_{i,\mathfrak L},\rho_{i,\mathfrak S}\in\mathbb Q^{\ge0}$ and
$\rho_{i,\mathfrak R}\in\mathbb Q^{\ge0}\cup\set\infty$. The
$\rho_{i,\mathfrak L}$ are the left bound, the $\rho_{i,\mathfrak R}$
are the right bound and the $\rho_{i,\mathfrak S}$ are the singletons.
Without loss of generality, it is assumed that the intervals are
disjoint and in increasing order.

Example \ref{ex:unbounded} is now resumed. 
\begin{example}\label{ex:decom-R-01}
  Let $R=\exRR$ as in Example \ref{ex:unbounded}.  Then $t_{R}$ is 4,
  $R_{0}=\left(\frac13,1\right]$, $R_{1}=\left[0,1\right]$, and
  $R_{2}=R_{3}=\set{0}\cup\left[\frac23,1\right]$. Furthermore, $I=3$,
  $J=2$, $\rho_{1,\mathfrak L}=\frac13$, $\rho_{2,\mathfrak R}=2$,
  $\rho_{2,\mathfrak{L}}=\frac83$, $\rho_{2,\mathfrak{R}}=3$,
  $\rho_{3,\mathfrak{L}}=\frac{11}3$, $\rho_{3,\mathfrak{R}}=\infty$,
  $\rho_{1,\mathfrak{S}}=2$ and $\rho_{2,\mathfrak{S}}=3$.
\end{example}

\section{Automata reading reals}\label{sec:aut-numb}
Automata recognizing encoding of set of reals are considered in this
section. The notion of Real Number Automata and of Fractional Number
Automata are introduced in Section \ref{sec:RNA-FNA}. Some sets of
states of the automata reading encoding of set of reals are considered
in Section \ref{sec:setOfState}.

\subsection{Real and Fractional Number Automata}\label{sec:RNA-FNA}
In this section, the automata reading saturated languages are
considered.

Following \cite{real}, a Büchi automaton of alphabet $\realDot$ is
said to be a Real Number Automaton (RNA) if \index{Real Number
  Automaton}\index{RNA- Real Number Automaton}\begin{itemize}
\item all words accepted by $\mathcal{A}$ contains exactly one $\realDot$,
  and
\item the language $\toInfWord{\mathcal A}$ is saturated.
\end{itemize}
The Büchi automata pictured in \ref{fig:ex-AR-R>0} and
\ref{fig:ex-AR-R>0-min} are RNA.  Clearly, the RNAs are the Büchi
automata which recognizes saturated languages of
$\digitSet^{*}\realDot\digitSet^{\omega}$.  Similarly, the name of
Fractional Number Automata (FNA) is given to the Büchi automata of
alphabet $\digitSet$ recognizing a saturated language.  \index{FNA -
  Fractional Number Automaton}\index{Fractional Number Automaton}

A weak {}Büchi automaton which is a RNA or a FNA is said to be a
\emph{weak RNA} or a \emph{weak FNA} respectively.  An example of FNA
is now given. This example shows that the number of intervals required
to describe a set is not polynomially bounded by the number of states
of the automaton recognizing this set.
\begin{example}\label{ex:A2-01}
  For every non-negative integer $n$, let $R_{n}$ be
  $\set{m2^{-n}\mid m\in[2^{n}]}$. It is the set of reals which admit
  an encoding $w$ in base $b$ whose suffixes $\suffix{w}{n}$ are either
  equal to $0^{\omega}$ or to $1^{\omega}$.  This set can not be
  described with less than $2^{n-1}$ intervals and is recognized by
  the automaton $\mathcal{A}_{n}$ with $n+3$ states:
  \begin{equation*}
    \mathcal{A}_{n}=\autPar{\set{q_{i}\mid i\in[n]}\cup\set{q_{n+1,0},q_{n+1,1},\emptyState}}{\digitSet}{\delta}{\iniState}{\set{q_{n+1,0},q_{n+1,1}}},
  \end{equation*}
  where the transition function is such that, for $a\in\digitSet[2]$:
  \begin{equation*}
    \arraycolsep=0.5pt
    \begin{array}{rclp{2mm}rclcrcl}
    \del{q_{i}}{a}&=&\multicolumn{5}{l}{q_{i+1}\mbox{ for }i\in[n-1],}&
    &\del{q_{n}}{a}&=&q_{n+1,a},\\
    \del{q_{n+1,a}}{a}&=&q_{n+1,a}&
    &\del{q_{n+1,a}}{1-a}&=&\emptyState&\mbox{ and }
    &\del{\emptyState}{a}&=&\emptyState.
    \end{array}
  \end{equation*}
  The automaton $\mathcal{A}_{3}$ is pictured in Figure \ref{fig:A2-01}, without
  the state $\emptyState$.
  
  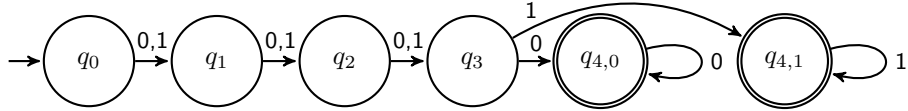
\begin{figure}[h]
    \centering
    \begin{tikzpicture}[->, >=stealth', shorten >=1pt, auto, node
      distance=1.7cm, thick, main node/.style={circle, fill=!20, draw,
        font=\sffamily\Large\bfseries, inner sep=0.1pt, minimum
        size=1cm}] \tikzset{every state/.style={minimum size=1.2cm}}
      \node[state, initial, initial text={}] (0) {$\iniState$};
      \node[state, right of=0] (1) {$q_{1}$};
      \node[state, right of=1 ] (2) {$q_{2}$};
      \node[state, right of=2 ] (3) {$q_{3}$};
      \node[state, right of=3,accepting ] (40) {$q_{4,0}$};
      \node[state, right=1.2cm of 40,accepting] (41) {$q_{4,1}$};

      \path[every node/.style={font=\sffamily\small}]
      (0)  edge              node {0,1} (1)
      (1)  edge       node {0,1} (2)
      (2)  edge       node {0,1} (3)
      (3)  edge        node {0} (40)
           edge [bend left]  node [above,pos=.08] {1} (41)
      (40) edge [loop right] node {0} (40)
      (41) edge [loop right] node {1} (41)
                 ;
    \end{tikzpicture}
    \caption{The automaton $\mathcal{A}_3$ of Example \ref{ex:A2-01},
      accepting
      $\set{0=\frac{0}{8},\frac{1}{8},\frac{2}{8},\frac{3}{8},\frac{4}{8},\frac{5}{8},\frac{6}{8},\frac{7}{8},\frac{8}{8}=1}$}
    \label{fig:A2-01}
  \end{figure}
\end{example}
\index{$\wordToReal{\mathcal{A}}$} For $\mathcal{A}$ a Büchi automaton of alphabet
$\digitSet$ (respectively, $\digitDotSet$). Let
$\wordToReal{\mathcal{A}}=\wordToReal{\toInfWord{\mathcal{A}}}$. It is a subset of
$[0,1]$ (respectively, $\mathbb R^{\ge0}$). It is said that $\mathcal{A}$ recognizes
$\wordToReal{\mathcal{A}}$.

It should be noted that two distinct minimal weak Büchi automata may
recognizes the same set of reals. Indeed, they may recognize two
distinct language which are two encoding of the same set of reals. At
least one of those languages is not saturated. Note however that two
distinct minimal RNA or RFA accepts distinct sets of reals.

\subsection{Some sets of states of RNA and of FNA}\label{sec:setOfState}
Five sets of states of Büchi automata are of used through this
paper. Those sets are introduced and studied in this section.

\begin{definition}[$\emptyStates$, $\zuStates$,
  $\inftyStates$, $\natStates$ and $\fraStates$]\label{not:q-empty-01}
  Let $\mathcal{A}$ be an automaton over alphabet $\digitDotSet$ or
  $\digitSet$.
  \begin{itemize}
  \item Let $\emptyStates$ \index{QO@$\emptyStates$} be the set of states
    $q$ such that $\mathcal{A}_{q}$ recognizes the empty language.
  \item Let $\zuStates$ \index{Q01@$\zuStates$} be the set of states
    $q$ such that $\mathcal{A}_{q}$ recognizes
    $\digitSet^{\omega}=\setRealToLanguage{[0,1]}$.
  \item Let $\inftyStates$ \index{Qinfty@$\inftyStates$} be the set of
    states $q$ such that $\mathcal{A}_{q}$ recognizes the language
    $\digitSet^{*}\realDot\digitSet^{\omega}=\setRealToLanguage{[0,\infty)}$.
  \item Let $\natStates$\index{Qnat@$\natStates$} be the set of states
    $q$ such that $\mathcal{A}_{q}$ recognizes a subset of
    $\digitSet^{*}\realDot\digitSet^{\omega}$.
      \item Let $\fraStates$\index{Qfra@$\fraStates$} be the set of states
    $q$ such that $\mathcal{A}_{q}$ recognizes a subset of
    $\digitSet^{\omega}$.
  \end{itemize}
\end{definition}
Example \ref{ex:unbounded-min} is now resumed.
\begin{example}
  Let $\mathcal{A}$ be the automaton pictured in
  Figure \ref{fig:ex-AR-R>0-min}. Let $\emptyState$ be the state
  $\del{(2,0)}{1}$, which is not pictured in
  Figure \ref{fig:ex-AR-R>0-min}.  Then $\zuStates=\set{\zuState[R]}$,
  $\inftyStates=\set{\inftyState[R]}$ and
  $\emptyStates=\set{\emptyState}$. Furthermore,
  $\natStates=\set{\iniState,
    q_{1},q_{2},\inftyState[R],\emptyState[R]}$,
  its elements are represented in the top row, of
  Figure \ref{fig:ex-AR-R>0-min}. Finally,
  $\fraStates=\set{(2,\epsilon), (2,0), (0,\epsilon), (0,0),
    \zuStates[R],\emptyState[R]}$.
  Its elements are pictured in the second row of
  Figure \ref{fig:ex-AR-R>0-min}.
\end{example}

The following lemma is straightforward from the definition.
\begin{lemma}\label{lem:min-empty-infty}
  In a minimal weak {}Büchi automaton, the
  sets $\emptyStates$, $\zuStates$ and $\inftyStates$ are either
  singletons or the empty set.
\end{lemma}
In a minimal weak {}Büchi automaton $\mathcal{A}$, let $\emptyState$,
\index{$\emptyState$} $\zuState$\index{$\zuState$} and
$\inftyState$\index{$\inftyState$} denote the only state $q$ such that
$\mathcal{A}_{q}$ recognizes the languages $\emptyset$,
$\digitSet^{\omega}$ and $\digitSet^{*}\realDot\digitSet^{\omega}$
respectively.  In an automaton of alphabet $\digitSet$, all states
belongs to $\fraStates$.

The following lemma states that those five sets are linear time
computable.
\begin{lemma}\label{lem:comp-set}
  Let $\mathcal{A}$ be a Büchi automaton{} with $n$ states{}. Then the sets $\emptyStates$, $\natStates$ and
  $\fraStates$ are computable in time $\bigO{n{b}}$. If
  $\mathcal{A}$ is weak{}, the sets $\zuStates$ and
  $\inftyStates$ are computable in time $\bigO{nb}$.
\end{lemma}
\begin{proof}
  Tarjan's algorithm \cite{Tarjan} can be used to compute the set of
  strongly connected component in time $\bigO{{nb}}$, and therefore
  the set of recurrent states. Furthermore, it is easy to associate in
  linear time to each state its set of predecessors. Let $p_{q}$ be
  the number of predecessors of a state $q$.
  
  Let us first explain how to compute the set $\emptyStates$. Note
  that $Q\setminus\emptyStates$ is the set of states $q$ such that
  $\mathcal{A}_{q}$ accepts some $\omega$-word. Hence
  $Q\setminus\emptyStates$ is the smallest set containing all
  accepting recurrent states and is closed under taking
  predecessors. Therefore $\emptyStates$ is the greatest set which
  does not contain the accepting recurrent states and is closed under
  taking successor. It can thus be computed by a fixed-point
  algorithm. The algorithm is now given.
  
  Two sets $S$ and $S'$ are used by the algorithm. The set $S$
  represents $\emptyStates$. The set $S'$ is the set of states of
  $\emptyStates$ which must be processed by the fixed-point algorithm.
  The algorithm initializes the set $S$ to $Q$ and initializes the set
  $S'$ to the empty set. The algorithm runs on each recurrent state
  $q$. For each state $q$, if $q$ is accepting, then $q$ is removed
  from $S$ and added to $S'$. The algorithm then runs on each element
  $q$ of $S'$. For each state $q$, the algorithms removes $q$ from
  $S'$ and runs on each predecessors $q'$ of $q$.  For each $q'$, if
  $q'$ is in $S$, then $q'$ is removed from $S$ and added to
  $S'$. Finally, when $S'$ is empty, the algorithm halts and
  $\emptyStates$ is the value of $S$.

  Let us now consider the computation time of this algorithm. At most
  $n$ states are added to $S'$, and each state is added at most
  once. For each state $q$ added to $S'$, each of its $c_{q}$
  predecessor is considered in constant time. Thus the algorithm runs
  in time $\bigO{n+ \sum_{q\in Q}c_{q}}=\bigO{nb}$.

  \paragraph{}It is now explained how to compute $\fraStates$ and
  $\natStates$.  Let $Q_{0}$, $Q_{1}$ and $Q_{2}$ be the set of states
  accepting a words with at least 0, 1 and 2 $\realDot$'s
  respectively.  Then $\fraStates$ is equal to
  $(Q_0\setminus{}Q_{1})\cup\emptyStates$ and $\natStates$ is equal to
  $(Q_{1}\setminus{}Q_{2})\cup\emptyStates$.  Let us now explain how
  to compute the sets $Q_{0}$, $Q_{1}$ and $Q_{2}$.  The set $Q_0$ is
  the smallest set containing all accepting recurrent states and
  closed under taking predecessors. The state $Q_{1}$ is the smallest
  set containing the predecessors of $Q_0$ by the letter $\realDot$
  and closed under taking predecessors. Similarly, the set $Q_{2}$ is
  the smallest set containing the predecessors of $Q_1$ by the letter
  $\realDot$ and closed under taking predecessors.  Those three sets
  are computable by a fixpoint algorithm similar to the one computing
  $\emptyStates$. It is thus computable in time $\bigO{{nb}}$.

  \paragraph{}
  Let us now assume that the Büchi automaton $\mathcal{A}$ is weak{}. It is now explained how to compute $\zuStates$
  and $\inftyStates$. Note that $\zuStates\subseteq\fraStates$ and
  that $\fraStates\setminus\zuStates$ is the set of states
  $q\in\fraStates$ such that there exists an $\omega$-word
  $w\in\digitSet^{\omega}$ which is not accepted by
  $\mathcal{A}_{q}$. Therefore, $\fraStates\setminus\zuStates$ is the
  smallest subset of $\fraStates$ containing non-accepting recurrent
  state and closed under taking predecessors by $\digitSet$.
  Similarly, note that $\inftyStates\subseteq\natStates$ and that
  $\natStates\subseteq\inftyStates$ is the set of states $q$ such that
  there is an infinite word of the form
  $\digitSet^{*}\realDot\digitSet^{\omega}$ which is not accepted by
  $\mathcal{A}_{q}$. Therefore $\natStates\subseteq\inftyStates$ is the
  smallest subset of $\natStates$ containing the predecessors of
  $\fraStates\setminus\zuStates$ by $\realDot$ and closed under taking
  predecessors by $\digitSet$.  The computation of $\natStates$ and of
  $\inftyStates$ is thus similar to the computation of $\emptyStates$.
\end{proof}
This lemma admits the following corollary.
\begin{corollary}\label{cor:correct-language}
  It is decidable in time $\bigO{nb}$ whether a Büchi automaton with
  $n$ states recognizes a subset of
  $\digitSet^{*}\realDot\digitSet^{\omega}$ or of
  $\digitSet^{\omega}$.
\end{corollary}
\begin{proof}
  By definition of $\natStates$ (respectively $\fraStates$), the
  automaton recognizes a subset of
  $\digitSet^{*}\realDot\digitSet^{\omega}$ (respectively of
  $\digitSet^{\omega}$) if and only if its initial state belongs to
  $\natStates$ (respectively $\fraStates$). By
  Lemma \ref{lem:comp-set}, it is testable in time $\bigO{bn}$.
\end{proof}

The following lemma gives a relation between the set of states
introduced in Example \ref{not:q-empty-01} and morphisms of automata.
\begin{lemma}\label{lem:morphism-sets}
  Let $\mathcal{A}=\aut$ and
  $\mathcal{A}'=\autPar{Q'}{\digitSet}{\delta'}{q'_{0}}{F'}$ be two Büchi
  automata. Let $\mu:Q\to Q'$ be a morphism of Büchi automaton. Then
  $\mu(\emptyStates)= \emptyStates[\mathcal{A}']$,
  $\mu(\fraStates)= \fraStates[\mathcal{A}']$,
  $\mu(\inftyStates)= \inftyStates[\mathcal{A}']$,
  $\mu(\natStates)= \natStates[\mathcal{A}']$, and
  $\mu(\zuStates)= \zuStates[\mathcal{A}']$.
\end{lemma}
\begin{proof}
  The proof is done for the first equality:
  $\mu(\emptyStates)= \emptyStates[\mathcal{A}']$. All other cases are
  similar. Let $q'\in Q'$, and let us prove that
  $q'\in\mu(\emptyStates)\iff{q'\in\emptyStates[\mathcal{A}']}$. It suffices
  to see that:
  \begin{equation*}
    \arraycolsep=0.9pt
    \begin{array}{rcllcl}
    q'\in\mu(\emptyStates)
    &\iff{}&\exists q\in Q.& q\in \emptyStates&\land&\mu(q)=q'\\
    &\iff{}&\exists q\in Q.& \toInfWord{\mathcal{A}_{q}}=\emptyset&\land&\mu(q)=q'\\
    &\iff{}&\exists q\in Q.& \toInfWord {\mathcal{A}'_{\mu(q)}}=\emptyset&\land&\mu(q)=q'\\
    &\iff{}&\exists q\in Q.& \toInfWord {\mathcal{A}'_{q'}}=\emptyset&\land&\mu(q)=q'\\
    &\iff{}&&\toInfWord{\mathcal{A}'_{q'}}=\emptyset\\
    &\iff{}&&q'\in\emptyStates[\mathcal{A}'].
    \end{array}
  \end{equation*}
\end{proof}
The following lemma gives a relation between the set of states
introduced in Example \ref{not:q-empty-01} and transitions. All of the
results follow easily from the definition of those sets.
\begin{lemma}
  Let $\mathcal{A}=\autPar{Q}{\digitDotSet}{\delta}{q_{0}}{F}$. Let $q\in Q$
  and $a\in\digitDotSet$. Then $\del{q}{a}$ belongs to the sets
  indicated in Example \ref{tab:Q-suc}.
  \begin{table}[h]
    \centering
    \begin{tabular}{|l|l|l|}
      \hline
      If $q$ belongs to & then $\del{q}{a}$, for $a\in\digitSet$, belongs to & and $\del{q}{\realDot}$ belongs to:\\
      \Xhline{4\arrayrulewidth}
      $\emptyStates$&$\emptyStates$&$\emptyStates$\\
      $\zuStates$&$\zuStates$&$\emptyStates$\\
      $\inftyStates$&$\inftyStates$&$\zuStates$\\
      $\natStates$&$\natStates$&$\fraStates$\\
      $\fraStates$&$\fraStates$&$\emptyStates$\\
      \hline
    \end{tabular}
    \caption{Set of states and transitions}\label{tab:Q-suc}
  \end{table}
\end{lemma}
\section{Three methods to prove decidability of
  automata problems}\label{sec:method}
In this section, three methods are given. Those methods allow to prove
that some problems over automata are decidable.

The method given in Section \ref{subsubsec:Honkala} is based on an
algorithm of \cite{Honkala86}, which decide whether an integer
automaton recognizes an ultimately periodic set of integer.  The
method given in Section \ref{subsubsec:muchnik} is based on an
algorithm of \cite{muchnik}, which decide whether an automaton reading
tuples of integers recognizes a set of tuples of integers which is
$\fo{\mathbb N;+}$-definable. Finally, the method given in Section
\ref{subsec:efficient} is based on an algorithm of \cite{Sakarovitch},
which decides in linear time whether a minimal automaton recognizes an
ultimately periodic set of integer.

Those algorithms are easily adapted to other decision problem on
automata reading vectors of integers or of reals.  While the methods
given in the first two sections are less efficient than the method of
the last section, it seems interesting to give those method as they
are very general and, as far as the author know, has not yet been
given in full generality. For example, it is shown that those methods
allow to prove that it is decidable:
\begin{itemize}
\item whether an automaton recognizes a set of real which is
  $\fo{\mathbb R;+,<,1}$-definable or $\fo{\mathbb R,\mathbb Z;+,<}$-definable
\item whether the recognized set is a subsemigroup of $(\mathbb R^{d},+)$.
\end{itemize}

Finally, the method given in Section \ref{subsec:efficient} is the more
efficient method of this section and is the method used in the
remaining of this paper. Note that this method leads to proofs which
are more complicated than the one needed to apply the two preceding
methods.

\subsection{Honkala and brute-force algorithm}\label{subsubsec:Honkala}
The method given in this section is based on \cite[Theorem
10]{Honkala86}.  Let $\mathcal L$ be a family of regular languages.
Conditions about $\mathcal L$ are now given. If a class of language
satisfies those condition, it is shown that it is decidable whether an
automaton recognizes a language belonging to $\mathcal L$.

Let us assume that there exists a size function $s:\mathcal L\to\mathbb N$, such
that:
\begin{enumerate}
\item \label{honk:s}the number of states of the minimal automaton
  recognizing a language $L\in\mathcal L$ is at least $s(L)$.
\item\label{honk:compute}a finite superset of $s^{-1}(i)$ is
  computable for all $i\in\mathbb N$.
\end{enumerate}

In order to decide whether a minimal automaton $\mathcal{A}$ with $n$ states
recognizes a language of $\mathcal L$, it suffices to consider the following
algorithm:
\begin{itemize}
\item The algorithm runs on each integer $i\in[n]$,
\item for each $i$, the algorithm runs on each language $L\in\mathcal L$
  such that $s(L)=i$, by Hypothesis \eqref{honk:compute} it can be done,
\item for each $L$, the algorithm constructs the minimal automaton
  $\mathcal{A}_{L}$ which recognizes $L$,
\item if the minimal automata of $\mathcal{A}$ and of $\mathcal{A}_{L}$ are equal,
  the algorithm accepts.
\end{itemize}
If the algorithm has not accepted, it rejects. 
\paragraph{}
The following proposition shows that this method can be applied to the
problem considered in this paper. Note that the algorithm given in the
following proof is inefficient.
\begin{proposition}\label{prop-honk-simple}
  It is decidable whether an automaton recognizes a
  $\fo{\mathbb R,\mathbb Z;+,<,1}$-definable set of reals.
\end{proposition}
The proposition also holds for the more general notion of automata
recognizing set of tuples of reals, as defined in
\cite{weak-R-+-vector}.
\begin{proof}
  Let $r\in \mathbb Q$ and $u.vw^{\omega}$ be one of its encoding in base $b$
  with $\length{v}$ and $\length{w}$ minimal.  Let the pre-periodic
  length of $r$ be $\length{u}+1+\length{v}$ and let the periodic
  length of $r$ be $\length{w}$.
  
  Let $\mathcal L$ be the class languages which are encodings of
  simple sets.  Let us first consider Hypothesis \ref{honk:s}.

  Let $R$ be a $\fo{\mathbb R,\mathbb Z;+,<,1}$-definable of real. By \cite[Section
  2]{elimination}, this logic admits quantifier elimination. Let
  $\phi\in\qf{\mathbb R,\mathbb Z;+,<,1}$ be a quantifier-free formula defining $R$.
  It can be proven that the number of state of the minimal automaton
  recognizing a simple set $R$ is, at least, the maximum of the
  pre-periodic lengths and of the periodic lengths of the constant
  appearing in $\phi$. Let $s(\phi)$ be the greatest pre-periodic or
  periodic length of the constants of $\phi$ and let $s(R)$ be the
  minimal $s(\phi)$ for any $\phi\in\qf{\mathbb R,\mathbb Z;+,<,1}$ which defines
  $R$.

  Let us now consider hypothesis \ref{honk:compute}, let us give an
  algorithm which takes as input an integer $i\in N$ and generates all
  sets $R$ such that $s(R)=i$. It suffices to remark that, for each
  $i\in\mathbb N$, there is a finite number of rationals whose pre-periodic
  length and whose periodic length is less than $i$, let $S^{\le i}$
  be the set of those rationals. The algorithm runs on each subset $U$
  of $S^{\le i}$. For each $U$ the algorithm generates all
  $\qf{\mathbb R,\mathbb Z;+,<,1}$-formulas in conjunctive normal form whose
  rationals belongs to $S^{\le i}$. Note that there is only a finite
  number of such formula up to permutation of the elements of the
  conjunctions. Those formulas define all of the sets $R$ such that
  $s(R)=i$.
\end{proof}
\subsection{Muchnik and decidable logic}\label{subsubsec:muchnik}
The algorithm given in this section is based on \cite[Theorem
3]{muchnik}.

Let $\mathfrak S$ be a set of subset of $(\mathbb R^{\ge0})^{d}$ be a set of
$d$-tuples of non-negative reals. Let $R$ be a $d$-ary symbol
representing a set of $d$-tuples of reals. Let us assume that there
exists a $\fo{\mathbb R,\mathbb Z;X_{b},+,<,1,R}$-formula $\phi$ which
characterizes whether a set $R$ belongs to $\mathfrak S$. It is
decidable whether a RVA $\mathcal{A}$ is such that
$\wordToReal{\mathcal{A}}\in\mathfrak S$. Indeed, the equivalence
between $\fo{\mathbb R,\mathbb Z;X_{b},+,<,1}$ and the Büchi automata is
effective, hence it suffices to translate the formula $\phi$ into a
Büchi automaton, according to the algorithm of \cite{weak-buchi-r-+},
where $R$ is encoded by the automaton $\mathcal{A}$.

Let us give an example of application of this method.
\begin{proposition}
  It is decidable whether a Büchi automaton recognizes a set which is a
  subsemigroup of $(\mathbb R^{d},+)$.
\end{proposition}
\begin{proof}
  Using the argument given above, it suffices to use the formula:
  \begin{eqnarray*}
    \forall x_{0},\dots,x_{d-1},
    y_{0},\dots,y_{d-1}. \left[ \left(x_{0},\dots,x_{d-1}\right)\in R\land
    \left(y_{0},\dots,y_{d-1}\right)\in R\right]\implies{}\\ {\left(x_{0}+y_{0},\dots,x_{d-1}+y_{d-1}\right)\in R}.
  \end{eqnarray*}
\end{proof}

Note that the method introduced in this section leads to inefficient
algorithm. The computation time of those algorithms are, at least, a
tower of exponential, whose height is equal to the number of
quantifier alternation.

\subsection{The efficient method}\label{subsec:efficient}

The method introduced in this section leads to proofs which are more
difficult than the method given in the two preceding
sections. However, this method also leads to more efficient
algorithms.  The method introduced in this section is the one used in
Theorems \ref{theo:aut-0-1}{ and} \ref{theo:aut-R}{}.  This method is
similar to the proofs used in \cite{Leroux} and in \cite{Sakarovitch}.

\begin{proposition}\label{prop:method}
  Let $\mathbb L'$ be a class of language and $\mathbb A'$ be a class of weak
  {}Büchi automata such that $L\in \mathbb L'$
  if and only if it is recognized by a Büchi automaton of $\mathbb A'$.
  
  Let $\mathbb L$ be a class of languages over an alphabet such that there
  exists a class $\mathbb A$ of weak {}Büchi
  automata such that:
  \begin{enumerate}
  \item\label{prop:method-algo} it is decidable in time $t(n,b)$
    whether a Büchi automaton belongs to $\mathbb A$, for $n$ the number
    of states and $b$ the number of letters,
  \item\label{prop:method-AR} for each $L\in \mathbb L\cap\mathbb L'$, there exists an
    automaton $\mathcal{A}\in\mathbb A$ which recognizes $L$,
  \item\label{prop:method-min} the minimal quotient of any automaton
    of $\mathbb A$ belongs to $\mathbb A$ and
  \item\label{prop:method-L} every language recognized by some automaton
    of $\mathbb A$ belongs to $\mathbb L$.
  \end{enumerate}
  There exists an algorithm $\alpha$, which halts in time
  $\bigO{t(n,b)}$, and decides whether a minimal automaton of $\mathbb A'$
  recognizes a language of $\mathbb L$. Furthermore, the algorithm
  $\alpha$ applied to an automaton belonging to
  $\mathbb A'\setminus \mathbb A$ may return a false negative but it may not
  return any false positive.
\end{proposition}
In this paper, $\mathbb A'$ is either the set of FNA or of
RNA. Considering the class $\mathbb A'$ allows to restrict the kind of
automata studied.  The proposition still hold when "weak
{}Büchi automata" is replaced by "finite
automaton". More generally, a similar proposition can be given as soon
as, for each language, there exists a canonical automaton recognizing
this language.  This requirement is the reason for which this
proposition does not hold for non-weak Büchi automata or for
non-deterministic automata.
\begin{proof}
  By Property \eqref{prop:method-algo} there exists an algorithm
  $\alpha$ which accepts in time $t(n,l)$ the automata of $\mathbb A$.
  Let $\mathcal{A}$ be a weak Büchi automaton and let
  $L=\toInfWord{\mathcal{A}}$. Let us prove that if $\alpha$ accepts $\mathcal{A}$
  then $L\in\mathbb L$ and if $L\in\mathbb L$ and $\mathcal{A}\in\mathbb A'$ then
  $\alpha$ accepts $\mathcal{A}$.

  Let us first prove that if $\alpha$ accepts $\mathcal{A}$, then
  $L\in \mathbb L$. If $\mathcal{A}$ is accepted by $\alpha$, by Property \eqref{prop:method-algo}, $\mathcal{A}\in\mathbb A$. By Property \eqref{prop:method-L}, since $\mathcal{A}\in\mathbb A$, $L\in\mathbb L$.

  Let us now prove that if $L\in\mathbb{L}$ and $\mathcal{A}\in\mathbb A'$ then
  $\alpha$ accepts $\mathcal{A}$.  Since $\mathcal{A}\in\mathbb A'$, by definition of
  $\mathbb{L'}$, $L\in\mathbb{L'}$.  Since $L\in\mathbb{L}$ and $L\in\mathbb{L'}$, by
  Property \eqref{prop:method-AR}, there exists an automaton
  $\mathcal{A}'\in\mathbb A$ which recognizes $L$.  Since $\mathcal{A}'$ and $\mathcal{A}$
  recognize the same languages and $\mathcal{A}$ is minimal, then $\mathcal{A}$ is
  the minimal quotient of $\mathcal{A}'$. Since $\mathcal{A}'$ is the minimal
  quotient of an automaton belonging to $\mathbb A$, by Property \eqref{prop:method-min} $\mathcal{A}\in\mathbb A$. Since $\mathcal{A}\in\mathbb A$, by
  Property \eqref{prop:method-algo}, the algorithm $\alpha$ accepts the
  automaton $\mathcal{A}$.
\end{proof}

\section{Automata accepting simple subsets of
  $[0,1]$}\label{sec:bounded}

This section deals with automata recognizing simple subsets of
$[0,1]$.  For each simple set $R$, a weak {}Büchi automaton which
recognizes $\toInfWord{R}$ is defined in Section
\ref{subsec:bounded-set->aut}. In Section
\ref{subsec:char:bounded-aut}, an algorithm is given, which accepts
the weak FNAs which recognize simple sets.

\subsection{From sets to automata}\label{subsec:bounded-set->aut}
Let $R\subseteq[0,1]$ be a simple set, in this section, it is
explained how to compute an automaton $\AR$ which recognizes
$\toInfWord{R}$. This automaton has the form descried in \cite[Chapter
5]{Boigelot2007}. As seen in Section \ref{sec:FU}, $R$ can be
expressed as:
\begin{equation*}
  \bigcup_{i=0}^{I-1}(\rho_{i,\mathfrak L},\rho_{i,\mathfrak R})\cup\bigcup_{i=0}^{J-1}\set{\rho_{i,\mathfrak S}}
\end{equation*}
with $\rho_{i,j}\in\mathbb Q$. In this section, it can furthermore be assumed
that the $\rho_{i,j}$'s belong to $[0,1]$.

Let $w_{i,j,k}$, with $k$ taking value in $\set{0,(b-1)}$, be the one
or two encodings in base $b$ of $\rho_{i,j}$. Let $l$ be an integer
such that, for all $i,j,k,i',j',k'$, either $w_{i,j,k}=w_{i',j',k'}$
or $\prefix{w_{i,j,k}}{l}\ne\prefix{w_{i',j',k'}}{l}$. By an easy
induction on the number of words $w_{i,j,k}$, such an integer $l$
exists.

Since the $\rho_{i,j}$ are rationals, their encodings in base $b$ are
of\index{u@$u_{i,j,k}$}\index{v@$v_{i,j,k}$} the form
$u_{i,j,k}v_{i,j,k}^{\omega}$ with $u_{i,j,k}\in\digitSet^{*}$,
$v_{i,j,k}\in\digitSet^{+}$ and $k\in\set{0,(b-1)}$. Without loss of
generality, let us assume that all $u_{i,j,k}> l$ and that
$\length{v_{i,j,k}}$ is minimal. Those two assumptions imply that no
$u_{i,j,k}$ is the prefix of a $u_{i',j',k'}$ and that if
$u_{i,j,k}=u_{i',j',k'}$ then $v_{i,j,k}=v_{i',j',k'}$.

A weak FNA $\AR$ which recognizes $\toInfWord{R}$ is now defined.
\begin{definition}[$\AR$]\index{AR@$\AR$ and $\AR'$ - for
    a simple set  $R\subseteq[0,1]$}
  \label{def:AR01}
  For $R$ a simple set, let
  $\AR=\autPar{Q_{R}}{\digitSet}{\delta_{R}}{q_{\epsilon}}{F_{R}}$
  where:
  \begin{itemize}
  \item The set of states $Q_{R}$ is defined as the set of states
    $q_{w}$ for $w$ a strict prefixes of $u_{i,j,k}v_{i,j,k}$, plus
    two states $\emptyState[R]$ and $\zuState[R]$.
  \item The transition function is
    \begin{equation*}
      \delta_{R}(q,a)=\left\{
        \begin{array}{ll}
          q&\mbox{ if }q\in\set{\emptyState[R],\zuState[R]},\\
          wa&\mbox{ if $q_{w}=q$ and $wa$ is a strict prefix of some $u_{i,j,k}v_{i,j,k}$}\\
          u_{i,j,k}&\mbox{ if $q=q_{w}$ and $wa=u_{i,j,k}v_{i,j,k}$},\\
          \zuState[R]&\mbox{ if $q=q_{w}$ and $\wordToReal{w a}$ belongs to an
                       interval $(\rho_{i,\mathfrak L},\rho_{i,\mathfrak R})$}\\
          \emptyState[R]&\mbox{ otherwise.}
        \end{array}
      \right.
    \end{equation*}
  \item The set $F_{R}$ of accepting states contains $\zuState[R]$ and
    the states $q_{w}$ for $w$ some strict prefix of
    $u_{i,\mathfrak S,k}v_{i,\mathfrak S,k}$ of length at least
    $\length{u_{i,\mathfrak S,k}}$, for some $i$ and $k$.
  \end{itemize}
\end{definition}
The last definition is used to characterize the automaton, not to
compute it, hence there is no need to study how to decide whether
$\wordToReal{wa}$ belongs to an interval
$(\rho_{i,\mathfrak L},\rho_{i,\mathfrak R})$, nor to compute the
words $u_{i,j,k}$. Note that the name of the states $\emptyState[R]$
and $\zuState[R]$ are consistent with Section \ref{sec:aut-numb}.

An automaton $\AR$ and its minimal quotient are now given. The set $R$
is not the same as the one  of Example \ref{ex:decom-R-01} because a
subset of $[0,1]$ is now required.
\begin{example}\label{ex:auto-R-01}
  Let $R=\exRfrac$ as in Example \ref{ex:decom-R-01}. The following table
  gives the values associated to those indexes.
  \begin{equation*}
      \begin{array}[c]{|llll|llll|}
        \hline
        \rho_{i,j}& \mbox{its  encodings in base 2}& u_{i,j,k}& v_{i,j,k}&
        \rho_{i,j}& \mbox{its  encodings in base 2}& u_{i,j,k}& v_{i,j,k}\\
        \Xhline{4\arrayrulewidth}
        \frac 14& 001^{\omega}& 001& 1&
        \frac13& (01)^{\omega}&0101&01\\
        \frac 14& 010^{\omega}& 0100& 0&
        \frac23& (10)^{\omega}&101&01\\
        \frac5{12}& 01(10)^{\omega}&011&01&
        &&&\\
        \hline
      \end{array}
    \end{equation*}

    The automaton $\AR$ is pictured in Figure \ref{fig:ex-AR-01}, without
    the state $\emptyState[R]$, and its minimal quotient is pictured in
    Figure \ref{fig:ex-AR-01}.
  \begin{figure}[!h]
    \centering
      \begin{tikzpicture}[->, >=stealth', shorten >=1pt, auto, node
        distance=1.7cm, thick, main node/.style={circle, fill=!20, draw,
          font=\sffamily\Large\bfseries, inner sep=0.1pt, minimum
          size=1cm}] \tikzset{every state/.style={minimum size=1.2cm}}
        
        \node[state,initial, initial text={}] (epsilon) {$q_{\epsilon}$};
        \node[state,  right of=epsilon]        (0)       {$q_{0}$};
        \node[state, below of=epsilon]        (1)       {$q_{1}$};
        \node[state, right=3cm  of 0]              (01)      {$q_{01}$};
        \node[state, below of=0]             (00)      {$q_{00}$};
        \node[state, accepting, below of=1]              (10)      {$q_{10}$};
        \node[state, accepting, below of=10]              (101)      {$q_{101}$};
        \node[state, accepting, right of=101]              (1010)      {$q_{1010}$};
        \node[state, accepting, right of=1010]              (10101)      {$q_{10101}$};
        \node[state, below of=00]              (001)     {$q_{001}$};
        \node[state, accepting, right of=001]              (0011)     {$q_{0011}$};
        \node[state, right of=01]              (010)     {$q_{010}$};
        \node[state,  below of=01]              (011)     {$q_{011}$};
        \node[state, right of=010]                       (0101)    {$q_{0101}$};
        \node[state, below of=011,accepting]             (0111)    {$q_{0111}$};
        \node[state, below of=0111,accepting]             (01110)    {$q_{01110}$};
        \node[state, below of=010,accepting]            (0100)    {$q_{0100}$};
        \node[state, right of=0101]                      (01010)   {$q_{01010}$};
        \node[state, below of=01010,accepting]           (f)       {$\zuState[R]$};
        
        \path[every node/.style={font=\sffamily\small}] 
        (epsilon) edge node  {0} (0)
        edge node  {1} (1)
        (0)    edge node  {0} (00)
               edge node  {1} (01)
        (1)    edge node  {0} (10)
        (00)   edge node  {1} (001)
        (01)   edge node  {0} (010)
               edge node  {1} (011)
        (10)   edge node  {1} (101)
        (101)   edge  node  {0} (1010)
        (1010)   edge [bend left]  node  {1} (10101)
        (10101)   edge [bend left]  node  {0} (1010)
        (001)  edge  node {1} (0011)
        (0011)  edge [loop above] node {1} (001)
        (010)  edge node  {0} (0100)
               edge node  {1} (0101)
        (011)  edge  node  {1} (0111)
        (0111) edge[bend left] node  {0} (01110)
        (01110) edge[bend left] node  {1} (0111)
        (0100) edge[loop below] node  {0} (0100)
               edge node  {1} (f)
        (0101) edge [bend left]node  {0} (01010)
        (01010)edge  node  {0} (f)
               edge [bend left] node {1} (0101)
        (f) edge [loop below] node {0,1} (f)
        ;
      \end{tikzpicture}
    \caption{The automaton $\AR$ for $R=\exRfrac$}
      \label{fig:ex-AR-01}
  \end{figure}
  \begin{figure}[!h]
    \centering
      \begin{tikzpicture}[->, >=stealth', shorten >=1pt, auto, node
        distance=1.7cm, thick, main node/.style={circle, fill=!20, draw,
          font=\sffamily\Large\bfseries, inner sep=0.1pt, minimum
          size=1cm}] \tikzset{every state/.style={minimum size=1.2cm}}

        \node[state,initial, initial text={}, below=.4cm of 10] (epsilon') {$q_{\epsilon}$};
        \node[state, right of=epsilon']        (0')       {$q_{0}$};
        \node[state, right=1.6cm  of 0']              (01')      {$q_{01}$};
        \node[state, below of=0', accepting]             (00')      {$q_{00}$};
        \node[state, below of=00', accepting]              (10')      {$q_{10}$};
        \node[state, left of=10', accepting]        (1')       {$q_{1}$};
        \node[state, right of=01', accepting]              (010')     {$q_{010}$};
        \node[state, right of=010']                       (0101')    {$q_{0101}$};
        \node[state, right of=0101']                      (01010')   {$q_{01010}$};
        \node[state, below of=0101', accepting]            (f')       {$\zuState[R]$};

        \path[every node/.style={font=\sffamily\small}] 
        (epsilon') edge node  {0} (0')
        edge node  {1} (1')
        (0')    edge node  {0} (00')
               edge node  {1} (01')
        (1')    edge [bend left] node  {0} (10')
        (00')   edge [loop right] node  {1} (00')
        (01')   edge node  {0} (010')
               edge [bend left] node  [below]{1} (10')
        (10')   edge [bend left] node  {1} (1')
        (010')  edge node  {0} (f')
               edge node  {1} (0101')
        (0101') edge [bend left] node {0} (01010')
        (01010')edge  node [below] {0} (f')
               edge [bend left] node {1} (0101')
        (f') edge [loop below] node {0,1} (f')
        ;
      \end{tikzpicture}
    \caption{The minimal  automaton recognizing $R=\exRfrac$}
      \label{fig:ex-AR-01-min}
  \end{figure}

  \paragraph{}
  Intuitively, the states $q_{1}$ and $q_{10}$ are used to read the
  binary encoding of $\frac23$. The states $q_{00}$ and $q_{001}$ are
  used to read one of the binary encoding of $\frac14$ and $q_{0100}$
  is used to read its other encoding. The states $q_{011}$ and
  $q_{0111}$ are used to read $\frac{11}{24}$. Finally, the states
  $q_{0101}$ and $q_{01010}$ are used to read $\frac13$.

  Note that minimization sends the states $q_{011}$ and $q_{10}$ to
  the same state, named $q_{10}$. Intuitively, it is because when an
  automaton reads a rational $r$ belonging to the boundary of $R$, the
  only important information is:
  \begin{itemize}
  \item the periodic part of the encoding of $r$,
  \item whether $r$ belongs to $R$ and
  \item whether, $(r-\epsilon,r)$ (respectively, $(r,r+\epsilon)$) is
    included in $R$ or is disjoint from $R$, for $\epsilon$ small
    enough.
  \end{itemize}
\end{example}

Let us prove that $\AR$ is as expected.
\begin{lemma}\label{lem:ar-01-correct}
  Let $R\subseteq[0,1]$ be a simple set. The automaton $\AR$ recognizes
  $\toInfWord{R}$.
\end{lemma}
\begin{proof}
  Let $w\in\digitSet^{\omega}$, and let $r=\wordToReal{w}$. Let us
  prove that $\wordToReal{w}\in R$ if and only if $w$ is accepted by
  $\AR$. Two cases must be considered, depending on whether $r$ is
  equal to some $\rho_{i,\mathfrak S}$ for some $i$, or not. In the
  second cases, four more cases must be considered, depending on
  whether $r$ is equal to some $\rho_{i,\mathfrak L}$, whether $r$ is
  equal to some $\rho_{i,\mathfrak R}$, whether $r$ belongs to some
  $(\rho_{i,\mathfrak L},\rho_{i,\mathfrak R})$ for some $i$, or
  whether $r$ does not satisfy any of those properties.
  \begin{itemize}
  \item Let us first assume that $r=\rho_{i,\mathfrak S}$ for some
    $i$. Then $r\in R$.  Let us show that $w$ is accepted by
    $\AR$. Since $r=\rho_{i,\mathfrak S}$, then $w$ is of the form
    $u_{i,\mathfrak S,k}v_{i,\mathfrak S,k}^{\omega}$, hence, by
    induction on the prefixes of $w$, all visited states are of the
    form $q_{w}$ for $w$ some prefix of
    $u_{i,\mathfrak S,k}v_{i,\mathfrak S,k}$. It implies that each
    state of the run of $\AR$ on $w$ -- apart from the
    $\length{u_{i,\mathfrak S,k}}$ first states -- are accepting.
    Hence $\AR$ accepts $w$.
  \item From now on, let us assume that, for all $i$,
    $r\ne\rho_{i,\mathfrak S}$.  Let us suppose that
    $r=\rho_{i,\mathcal L}$ for some $i$. The case $r=\rho_{i,\mathcal R}$ is
    similar.  In this case, $r\not\in R$. Let us prove that $w$ is not
    accepted by $\AR$. The word $w$ is of the form
    $u_{i,j,k}v_{i,j,k}^{\omega}$, hence by induction on the prefix's
    of $w$, each visited state is $q_{w}$ for $w$ a prefix of
    $u_{i,j,k}v_{i,j,k}$, and since the $u_{i,j,k}$ are not prefixes
    of $u_{i',j',k'}v_{i',j',k'}^{\omega}$, then no accepting state is
    visited.  Hence $\AR$ does not accepts $w$.
  \item Let us assume that
    $r\in(\rho_{i,\mathfrak L},\rho_{i,\mathfrak R})$. Let us show
    that $\AR$ accepts $w$. Three cases must be considered, depending
    on whether $u_{i,\mathfrak L,k}$ is a prefix of $w$ for some
    $k\in\set{0,(b-1)}$, whether $u_{i,\mathfrak R,k}$ is a prefix or
    $w$ for some $k\in\set{0,(b-1)}$, or whether none of those words
    are prefix of $w$.
    \begin{itemize}
    \item Let us suppose that $u_{i,\mathfrak L,k}$ is a prefix of
      $w$, then there exists a unique 4-tuple $n\in \mathbb N$,
      $h<\length{v_{i,\mathfrak L,k}}$, $a<v_{i,\mathfrak L,k}[h]$ and
      $w'\in\digitSet^{\omega}$ such that
      \begin{equation*}
        w=u_{i,\mathfrak L,k}v_{i,\mathfrak L,k}^{n}(\prefix{v_{i,\mathfrak L,k}}{h})aw'.
      \end{equation*}
      Since $a<v_{i,\mathfrak L,k}[h]$,
      $\wordToReal{u_{i,\mathfrak L,k}\prefix{v_{i,\mathfrak
            L,k}}{h}aw'}>\rho_{i,\mathfrak L}$.
      Since the $u_{i,j,k}$'s are not prefixes of the
      $u_{i',j',k'}$'s,
      $\wordToReal{u_{i,\mathfrak L,k}\prefix{v_{i,\mathfrak L,k}}{h}aw'}<\rho_{i,\mathfrak L}$. Then, for
      $i>\length{u_{i,\mathfrak L,k}v_{i,\mathfrak L,k}^{n}\prefix{v_{i,\mathfrak
            L,k}}{h}}$,
      the $i$th state is $\zuState[R]$, hence $w$ is accepted by $R$.
    \item If $u_{i,\mathfrak R,k}$ is a prefix of $w$, the proof is
      similar by symmetry.
    \item Let us now assume that there are no $j$ and $k$ such that
      $u_{i,j,k}$ is a prefix of $w$. Let us consider the longest
      prefix $p$ of $w$ and $u_{i,j,k}$.  The $\length{p}$ first steps
      of the run of $\AR$ on $w$ are the states $q_{v}$ for $v$ some
      prefix of $p$. The following steps are $\zuState[R]$, hence
      $\AR$ accepts $w$.
    \end{itemize}
  \item In all other cases, the proof is similar to the preceding
    case, replacing $\zuState[R]$ by $\emptyState[R]$.
  \end{itemize}
\end{proof}

\subsection{Characterization of automata recognizing simple
  sets}\label{subsec:char:bounded-aut}
The main theorem of this paper concerning subsets of $[0,1]$ is now
stated.
\begin{theorem}\label{theo:aut-0-1}
  It is decidable in time $\bigO{nb}$ whether a minimal FNA over the
  alphabet $\digitSet$ with $n$ states recognizes a simple set.
\end{theorem}

The proof of this theorem uses Proposition \ref{prop:method}.  In order to use
this proposition, a set $\BAF$ of automata is now introduced. Four
lemmas are then proved, which corresponds to the four properties of
Proposition \ref{prop:method}.
\begin{definition}[$\BAF$]\index{A@$\BAF$}\label{def:BAF}
  Let $\BAF$ be the set of weak {}Büchi
  automata $\mathcal{A}=(Q,\digitSet,\delta,\iniState,F)$ such that, for each
  strongly connected component
  $C\subseteq \fraStates\setminus(\zuStates\cup{}\emptyStates)$, there
  exists $\beta_{<,C}$ and $\beta_{>,C}$, two states of
  $\zuStates\cup{}\emptyStates$, such that, for all $q\in C$:
  \begin{enumerate}
  \item\label{def:BAF-cycle} $C$ is a
    cycle,
  \item\label{def:BAF-cycle>} for all $a>a_{q}$, $\delta(q,a)$ is
    $\beta_{>,C}$ and
  \item\label{def:BAF-cycle<} for all $a<a_{q}$, $\delta(q,a)$ is
    $\beta_{<,C}$.
  \end{enumerate}
\end{definition}

Property \eqref{def:BAF-cycle>} implies that $\zuStates\cup\emptyStates$
is not empty.

Example \ref{ex:auto-R-01} is resumed in order to show the
construction of the preceding lemma.
\begin{example}\label{ex:auto-R-01-aut}
  Let $R=\exRR
  $
  and $\mathcal{A}$ be the minimal automaton of Figure
  \ref{fig:ex-AR-01-min}. Let us first consider the recurrent states
  $q$ equal to $1$, $00$, $10$, $0101$ or to $01010$. The integer $n$,
  the sequence of letters $a_{0},\dots,a_{n-1}$, the states
  $\beta_{<,C}$ and $\beta_{>,C}$ associated to each of those states
  $q$ are given in the following table and the Boolean $\beta_{=,C}$
  which is true if and only if $C$ is composed of accepting states.
  The two last columns of the table show the language of infinite
  words recognized by the automaton $\mathcal{A}_{q}$, and the set of
  reals recognized by this state.
  \begin{equation*}
    \begin{array}[c]{lllllllll}
      \hline
      q& n& a_{0}&a_{1}&\beta_{<,C}&\beta_{=,C}&\beta_{>,C}& \toInfWord{\mathcal{A}_{q}}& \wordToReal{\mathcal{A}_{q}}\\
      \Xhline{4\arrayrulewidth}
      1& 2& 0   & 1   & \emptyState  & \true  & \emptyState   & (01)^{\omega}&\left\{\frac13\right\}\\
      10& 2& 1   & 0   & \emptyState  & \true  & \emptyState   & (10)^{\omega}&\left\{\frac23\right\}\\
      00& 1& 1   &    & \emptyState  & \true  & \emptyState   & (1)^{\omega}&\left\{1\right\}\\
      0101& 2& 0   & 1   & \zuState  & \false  & \emptyState   & 0(10)^{\omega}+(10)^{*}0(0+1)^{\omega}&\left[0,\frac13\right)\\
      01010& 2& 1   & 0   & \zuState  & \false  & \emptyState   & (10)^{\omega}+(10)^{*}0(0+1)^{\omega}&\left[0,\frac23\right)\\
      \zuState[R]&&&& &&& (0+1)^{\omega}&\left[0,1\right]\\
      \emptyState[R]&&&& &&& \emptyset&\emptyset\\
      \hline
    \end{array}
  \end{equation*}
  \paragraph{}
  Transient states $010$, $01$, $0$ and $\epsilon$ are now
  considered.  One has:
  \begin{equation*}
    \arraycolsep=0.5pt
    \begin{array}{ccccccccccccccc}
      \wordToReal{\mathcal{A}_{010}}     &=&\frac{0+\wordToReal{\mathcal{A}_{\zuState[R]}}}{2}&\cup&\frac{1+\wordToReal{\mathcal{A}_{0101}}}{2}&=&\frac{0+\left[0,1\right]}{2}                                   &\cup&\frac{1+\left[0,\frac13\right)}{2}                                
      &=&\left[0,\frac23\right),\\
      \wordToReal{\mathcal{A}_{01}}      &=&\frac{0+\wordToReal{\mathcal{A}_{010}}}{2}        &\cup&\frac{1+\wordToReal{\mathcal{A}_{10}}}{2}  &=&\frac{0+\left[0,\frac23\right)}{2}                             &\cup&\frac{1+\set{\frac23}}{2}                                         &
                                                                                                                                                                                                                                                                                                                                                                                                                             =&\left[0,\frac13\right)\cup\set{\frac56},\\
      \wordToReal{\mathcal{A}_{0}}       &=&\frac{0+\wordToReal{\mathcal{A}_{00}}}{2}         &\cup&\frac{1+\wordToReal{\mathcal{A}_{01}}}{2}  &=&\frac{0+\left\{1\right\}}{2}                                   &\cup&\frac{1+\left\{\left[0,\frac13\right)\cup\set{\frac56}\right\}}{2}&=
        &\left[\frac12,\frac23\right)\cup\set{\frac{11}{12}},\\
      \wordToReal{\mathcal{A}_{\epsilon}}&=&\frac{0+\wordToReal{\mathcal{A}_{0}}}{2}          &\cup&\frac{1+\wordToReal{\mathcal{A}_{1}}}{2}   &=&\frac{0+\left[\frac12,\frac23\right)\cup\set{\frac{11}{12}}}{2}&\cup&\frac{1+\set{\frac13}}{2}                                         
      &=&\exRR
.
    \end{array}
  \end{equation*}
\end{example}
It is now proven that the set $\BAF$ satisfies Property \eqref{prop:method-algo} of Proposition \ref{prop:method}.
\begin{lemma}\label{lem:dans-baf}
  It is decidable in time $\bigO{nb}$ whether a weak {}Büchi automaton
  with $n$ states belongs to $\BAF$.
\end{lemma}
\begin{proof}
  Tarjan's algorithm \cite{Tarjan} can be used to compute the set of
  strongly connected component in time $\bigO{nb}$ and thus the set of
  recurrent state. By Lemma \ref{lem:comp-set}, the sets $\emptyStates$
  and $\zuStates$ are computable in linear time. The algorithm runs on
  each strongly connected component distinct from $\zuStates$ and from
  $\emptyStates$.

  It is now explained how the algorithm checks whether Property
  \eqref{def:BAF-cycle} is satisfied by the automaton. The algorithm
  runs on each $q\in Q$. The algorithm keeps a counter $c_{q}$,
  initialized to 0, of the number of letters $a\in\digitSet$ such that
  $\del{q}{a}\in C$.  For each $q$, the algorithm runs on each letter
  $a\in\digitSet$.  For each $a$, the algorithm tests whether
  $\del{q}{a}\in C$, and if it is the cases, $c_{q}$ is
  incremented. If $c_{q}\ne 1$ the algorithm rejects.
  
  It is now explained how the algorithm checks whether Property
  \eqref{def:BAF-cycle>} is satisfied by the automaton. Checking
  Property \eqref{def:BAF-cycle<} is done similarly. The algorithm
  runs on each $q\in C$. If $a_{q}>0$ then:
  \begin{itemize}
  \item if $\beta_{<,C}$ is not set, then $\beta_{<,C}$ is set to
    $\del{q}{0}$.
  \item otherwise, let us assume that $\beta_{<,C}$ is set.  The
    algorithm runs on each $0\le a<a_{q}$. For each $a$, if
    $\del{q}{a}$ is different from $\beta_{<,C}$, then the algorithm
    rejects.
  \end{itemize}

  If the algorithm has not rejected, it accepts.
\end{proof}
It is now proven that the set $\BAF$ satisfies Property \eqref{prop:method-AR} of Proposition \ref{prop:method}.
\begin{lemma}\label{lem:AR-in-BAF}
  Let $R\subseteq [0,1]$ be a simple set. The automaton $\AR$ belongs
  to $\BAF$.
\end{lemma}
\begin{proof}
  It suffices to see that the recurrent states
  $q\not\in\emptyStates\cup \zuStates$ of $\AR$ are of the form
  $u_{i,j,k}w$ with $w$ a prefix of $v_{i,j,k}$. For each prefix $w$
  of length $l$ of $v_{i,j,k}$, the digit $\suc{q_{u_{i,j,k}w}}$ is
  $v_{i,j,k}[l]$, thus Property \eqref{def:BAF-cycle} holds. The state
  $\beta_{<,C}$ (respectively, $\beta_{>,C}$) is $\emptyState$ if
  $u_{i,j,k}0^{\omega}\not\in R$ and $\zuState$ otherwise, thus
  Property \eqref{def:BAF-cycle>} (respectively \ref{def:BAF-cycle<})
  holds.  Then the conditions of Definition \ref{def:BAF} are
  satisfied.
\end{proof}

It is now proven that the set $\BAF$ satisfies  Property \eqref{prop:method-min} of Proposition \ref{prop:method}. 
\begin{lemma}\label{lem:quotient-BAF}
  The minimal quotients of automata of $\BAF$ belong to $\BAF$.
\end{lemma}
\begin{proof}
  Let $\mathcal{A}=\autPar{Q}{\digitSet}{\delta}{\iniState}{F}$ be an
  automaton belonging to $\BAF$ and let its minimal quotient be
  $\mathcal{A}'=\autPar{Q'}{\digitSet}{\delta'}{q'_{0}}{F'}$. Let
  $\mu$ be the morphism from $\mathcal{A}$ to $\mathcal{A}'$. Let us
  show that $\mathcal{A}'$ belongs to $\BAF$.  Let $C'$ be a component
  included in $\mathcal{A}'$, distinct from
  $\emptyStates[\mathcal{A}']$ and from $\zuStates[\mathcal{A}']$. By
  Lemma \ref{lem:scc-morphism-codomain}, there exists a strongly
  connected component $C$ of $\mathcal{A}$ such that $\mu(C)=C'$ and
  such that, for all $q\in Q\setminus C$ accessible from $C$,
  $\mu(q)\not\in C'$.

  Let us first show that Property \eqref{def:BAF-cycle} is satisfied by
  $\mathcal{A}'$. Let $q'\in C'$ and let us prove that there exists
  exactly one letter $\suc{q'}$ such that $\dell{q'}{\suc{q'}}\in C'$.
  Since $q'$ is recurrent, at least one such letter exists. It remains
  to prove the unicity. Let us assume that there exists two letters
  $a_{0}$ and $a_{1}$ such that $\dell{q'}{a_{0}}\in C'$ and
  $\dell{q'}{a_{1}}\in C'$ and let us prove that $a_{0}=a_{1}$.  Since
  $\mu(C)=C'$, there exists $q\in C$ such that $\mu(q)=q'$. Since
  $\mathcal{A}\in\BAF$, there exists exactly one letter $\suc{q}$ such
  that $\del{q}{\suc{q}}\in C$.  It suffices to prove that
  $a_{0}=\suc{q}=a_{1}$. Let us prove that $a_{0}=\suc{q}$, the other
  case is similar. Since
  $\mu(\del{q}{a_{0}})=\dell{\mu(q)}{a_{0}}=\dell{q'}{a_{0}}\in C'$
  and since $\del{q}{a_{0}}$ is accessible from $C$, by hypothesis
  about $C$, $\del{q}{a_{0}}\in C$. Since $\del{q}{a_{0}}\in C$, by
  Property \eqref{def:BAF-cycle} applied to $\mathcal A$, by definition of
  $\suc{q}$, $a_{0}=\suc{q}$.

  It is now proven that $\mathcal{A}'$ satisfies Property \eqref{def:BAF-cycle>}. The case of Property \eqref{def:BAF-cycle<} is
  similar. Let $q'_{1}$ and $q'_{2}$ be two states of $C'$, and let
  $a_{1}>\suc{q'_{1}}$ and $a_{2}>\suc{q'_{2}}$. Since $\mu(C)=C'$,
  there exists $q_{1},q_{2}\in C$ such that $\mu(q_{1})=q'_{1}$ and
  $\mu(q_{2})=q'_{2}$.  Let $q_{1}$ and $q_{2}$ be those two
  antecedents, note that $\suc{q_{1}}=\suc{q'_{1}}$ and
  $\suc{q_{2}}=\suc{q'_{2}}$. Since $\mathcal{A}\in\BAF$, by Property applied to
  $\mathcal A$ \ref{def:BAF-cycle>}
  $\del{q_{1}}{a_{1}}=\del{q_{2}}{a_{2}}$, hence
  $\dell{q'_{1}}{a_{1}}=\dell{q'_{2}}{a_{2}}$.
\end{proof}
It is now proven that the set $\BAF$ satisfies Property \eqref{prop:method-L} of Proposition \ref{prop:method}.
\begin{lemma}\label{lem:BAF->pour}
  The automata of $\BAF$ recognize simple subsets of $[0,1]$.
\end{lemma}
Note that it is not required in this lemma that the Büchi automaton is
a FNA. In order to prove this lemma, another lemma is now
required. Its proof is straightforward from Definition \ref{def:BAF}.
\begin{lemma}\label{lem:BAF->pour:aux}
  Let $\mathcal{A}$ be a Büchi automaton of $\BAF$ and let $q$ be a state of
  $\mathcal{A}$. The automaton $\mathcal{A}_{q}$ belongs to $\BAF$.
\end{lemma}

\begin{proof}[Proof of Lemma \ref{lem:BAF->pour}]
  The proof is by induction on the number $n$ of states of
  $\mathcal{A}$.  Two cases must be considered, depending on whether
  the initial state of the automaton is recurrent or not.

  Let us first suppose that the initial state is transient.  For each
  $a\in\digitSet$, by Lemma \ref{lem:BAF->pour:aux}, the automaton
  $\mathcal{A}_{\delta(\iniState,a)}$ belongs to $\BAF$.  The automaton
  $\mathcal{A}_{\delta(\iniState,a)}$ has less than $n$ states and belongs to
  $\BAF$, hence, by induction hypothesis, it recognizes a set $R_{a}$ of
  the form
  $\bigcup_{i}^{I_a-1}(\rho_{i,\mathfrak L,a},\rho_{i,\mathfrak R,a})\cup\bigcup_{i=0}^{J_a-1}\set{\rho_{i,\mathfrak S,a}}$. The
  set $\wordToReal{\mathcal{A}}$ is then equal to:
  \begin{equation*}
    \bigcup_{a=0}^{b-1}\frac{a+R_{a}}{b}\\
    =\bigcup_{a=0}^{b-1}\bigcup_{i=0}^{I_a-1}\left(\frac{a+\rho_{i,\mathfrak L,a}}b,\frac{a+\rho_{i,\mathfrak R,a}}b\right)\cup\bigcup_{i=0}^{J_a-1}\left\{\frac{a+\rho_{i,\mathfrak S,a}}b\right\},
  \end{equation*}
  where the $\rho_{i,j}$ belongs to $\mathbb Q\cap[0,1]$. Hence
  $\wordToReal{\mathcal{A}}$ is a simple set.
  \paragraph{}
  Let us now assume that the initial state is recurrent. Three cases
  must be considered depending on whether $\iniState$ belongs to
  $\emptyStates$, to $\zuStates$ or to neither of those sets.  Let us
  first assume that $\iniState\in{}\emptyStates$, the case
  $\iniState\in{}\inftyStates$ is similar.  The automaton recognizes
  $\emptyset$ by definition of $\emptyStates$, thus is simple.

  Let us suppose that $\iniState\not\in\emptyStates\cup\zuStates$ and
  let $C$ be the strongly connected component of $\iniState$. By
  Property \eqref{def:BAF-cycle} of Definition \ref{def:BAF}, there
  exists a sequence $a_{0},\dots,a_{n-1}$ such that
  $\del{\iniState}{a_{0}\dots a_{n-1}}=\iniState$.  Let
  $c_{\iniState}$ be the real represented by
  $\left(a_{0}\dots a_{n'-1}\right)^{\omega}$ that is
  $\frac{\sum_{i=0}^{n'-1}a_{i}b^{n'-i-1}}{b^{n'}-1}$. The number
  $c_{\iniState}$ is rational, and its length is
  $\bigO{\log(b)n}$. Let $w\in\digitSet^{\omega}$ and
  $x=\wordToReal{w}$. It is then clear from Definition \ref{def:BAF},
  that $\mathcal{A}$ accepts $w$ if and only if, either $x=c_{q_{0}}$
  and $C\subseteq F$, either $x<c_{q_{0}}$ and
  $\beta_{<,C} \in \fraStates$, or similarly $x>c_{q_{0}}$ and
  $\beta_{>,C} \in \fraStates$.
\end{proof}

Theorem \ref{theo:aut-0-1} can now be proven.
\begin{proof}
  It suffices to use Proposition \ref{prop:method} with $\mathbb A$ being the set
  of automata $\BAF$, $\mathbb A'$ being the set of FNA, and Lemmas
  \ref{lem:dans-baf}, \ref{lem:AR-in-BAF}, \ref{lem:quotient-BAF} and
  \ref{lem:BAF->pour}.
\end{proof}
Note that the algorithm given in the proof of Theorem \ref{theo:aut-0-1}
returns no false positive even when it is applied to a Büchi Automaton
which is not a FNA.  The author conjecture that there exists no false
negative.  

\section{Simple subsets of $\mathbb R^{\ge0}$}\label{sec:real-aut}
In the preceding section, the problem studied in this paper was solved
on $[0,1]$. The general problem is solved in this section using the
notations and lemmas of Section \ref{sec:bounded}.

In Section \ref{subsec:set->aut}, given a simple set $R$, a weak
{}automaton $\AR$ is constructed, which
recognizes $R$. In Section \ref{subsec:char:aut}, an algorithm is given,
which takes as input a weak {}automaton of
alphabet $\digitDotSet$ and decides whether it recognizes a simple
set.

\subsection{From sets to automata}\label{subsec:set->aut}
Let us fix in this section a simple set $R\subseteq\mathbb R^{\ge0}$. Since
$R$ is a simple set, there exists a least integer $t_{R}\in\mathbb N^{\ge0}$
such that $[t_{R},\infty)$ is either a subset of $R$ or is disjoint
from $R$.  For all $i\in\mathbb N$, let $R_{i}$ denote
$\set{x\in[0,1]\mid x+i\in R}$, and let
$\ARPar{{i}}=\autPar{Q_{i}}{\digitSet}{\delta_{i}}{q_{0,i}}{F_{i}}$
be the minimal automaton accepting $\wordToReal{R_{i}}$. By
Example \ref{ex:unbounded} is now resumed.
\begin{example}\label{ex:unbounded-follow}
  Let $R=\exRR$ as in Example \ref{ex:unbounded}. Then $t_{R}$ is 4,
  $R_{0}=\left(\frac13,1\right]$, $R_{1}=\left[0,1\right]$, and
  $R_{2}=R_{3}=\set{0}\cup\left[\frac23,1\right]$. 
\end{example}

In this section, a weak RNA $\AR$ which recognizes $\toInfWord{R}$ is
constructed. The part of $\AR$ which reads the fractional parts of the
reals is based on the construction of
Section \ref{subsec:bounded-set->aut}.  The formal definition of $\AR$ is
now given.
\begin{definition}[$\AR$]\index{AR@$\AR$ - for
    $R\subseteq\mathbb R^{\ge0}$}\label{def:AR-FU}
  Let $R\subsetneq[0,\infty)$ be a simple non-empty set. Note that
  $t_{R}>0$.  Let $\AR$ be the automaton
  $(Q,\digitDotSet,\delta,\iniState,F)$ where:
  \begin{itemize}
  \item the set $Q$ of state contains:
    \begin{itemize}
    \item a state $q_{i}$ for all $i\in[t_{R}-1]$,
    \item a state $\emptyState[R]$,
    \item a state $\inftyState[R]$ if $[T_{R},\infty)\subseteq R$,
    \item a state $(i,q)$ for all $i\in[t_{R}-1]$, and for each state
      $q\in Q_{i}\setminus \emptyState[\mathcal{A}_{i}]$.
    \end{itemize}
  \item The initial state is $q_{0}$.
  \item The accepting states are $\zuState$ and $(i,q)$ for all
    accepting states $q$ of $\mathcal{A}_{i}$.
  \item The transition function is such that, for all $i\in[t_{R}-1]$,
    $a\in\digitSet$:
    \begin{itemize}
    \item $\delta(q,a)=q$ for $q$ being $\zuState$, $\inftyState[R]$
      or $\emptyState[R]$,
    \item $\delta(q,\realDot)=\emptyState[R]$ for $q$ being $\zuState$ or
      $\emptyState[R]$,
    \item $\delta(\inftyState[R],\realDot)$ is $\zuState[R]$,
    \item $\delta(q_{i},a)$ is $bi+a$ if $bi+a< t_{R}$,
    \item $\delta(q_{i},a)$ is  $\inftyState[R]$  if $bi+a\ge t_{R}$
      and if $[t_{R},\infty)\subseteq R$,
    \item $\delta(q_{i},a)$ is $\emptyState[R]s$ if $bi+a\ge t_{R}$
      and if $[t_{R},\infty)\cap R=\emptyset$,
    \item $\delta(q_{i},\realDot)=(i,q_{0,i})$,
    \item $\delta((i,q),a)$ is $(i,\delta_{i}(q,a))$ for
      $q\in\mathcal{A}_{i}$, if $\delta_{i}(q,a)\not\in\emptyStates[\mathcal{A}_{i}]$ 
    \item $\delta((i,q),a)$ is $\emptyState$ for $q\in\mathcal{A}_{i}$, if
      $\delta_{i}(q,a)\in\emptyStates[\mathcal{A}_{i}]$ and
    \item $\delta((i,q),\realDot)$ is $\emptyState[R]$.
    \end{itemize}
  \end{itemize}
\end{definition}
An example of automaton $\mathcal{A}_{R}$ is now given, resuming
Example \ref{ex:unbounded-follow}.
\begin{example}\label{ex:construct-RR}
  Let $R=\exRR$, as in Example \ref{ex:unbounded-follow}.  The automaton
  $\AR$ is pictured in Figure \ref{fig:ex-AR-R>0}, without the non
  accepting state $\emptyState$. Its minimal quotient is pictured in
  Figure \ref{fig:ex-AR-R>0-min}.
\end{example}

Let us now show that $\AR$ is as expected.
\begin{proposition}\label{prop:ar-recognizes-R-unbound}
  Let $R\subsetneq\mathbb R^{\ge0}$ be a simple non-empty set. The automaton
  $\AR$ recognizes $\toInfWord{R}$.
\end{proposition}
\begin{proof}
  Let $\natPart{w}\in\digitSet^{*}$, $\fraPart{w}\in\digitSet^{\omega}$ and
  $w=\natPart{w}\realDot{}\fraPart{w}$. Let $\natPart{r}=\wordToNatural{\natPart{w}}$,
  $\fraPart{r}=\wordToReal{\fraPart{w}}$ and $r=\wordToReal{w}=\natPart{r}+\fraPart{r}$.  Let us prove that
  $\wordToReal{w}\in R$ if and only if $w$ is accepted by $\AR$.
  
  By an easy induction on the length of $\natPart{w}$,
  $\del{\iniState}{\natPart{w}}$ is $q_{\natPart{w}}$ if
  $\natPart{r}<t_{R}$, otherwise it is $\inftyState[R]$ if
  $[t_{R},\infty)\subseteq R$ and it is $\emptyState$ otherwise. Two
  cases are considered, depending on whether $\natPart{r}<t_{R}$ or
  whether $\natPart{r}\ge t_{R}$.
  \begin{itemize}
  \item Let us first assume that $\natPart{r}<t_{R}$. Then
    $\delta(0,\natPart{w})=q_{\natPart{w}}$ and thus
    $\delta(0,\natPart{w}\realDot{})=(\natPart{r},q_{0,\natPart{r}})$. By
    Lemma \ref{lem:ar-01-correct}, $\mathcal{A}_{\natPart{r}}$ recognizes
    $R_{\natPart{r}}$, hence $\fraPart{w}$ is accepted by
    $(\AR)_{(\natPart{r},q_{0,\natPart{r}})}$ if and only if
    $\fraPart{r}\in \ARPar{{\natPart{r}}}$. Hence $w$ is accepted by $\AR$
    if and only if $r\in R$.
  \item Let us now assume that $\natPart{r}\ge t_{R}$.  Let us assume
    that $[t_{R},\infty)\subseteq R$, the case
    $[t_{R},\infty)\cap R=\emptyset$ is similar.  Since
    $r_{i}\ge t_{R}$, then $r\in R$. By definition of $\AR$,
    $\del{\iniState}{\natPart{w}}=\inftyState[R]$, therefore
    $\del{\iniState}{\natPart{w}\realDot}=\zuState$. It follows that
    each state of the run of $\AR$ on $w$ is $\zuState$, apart from
    the $\length{\natPart{w}\realDot}$ first ones. Since furthermore
    $\zuState$ is an accepting state, $\AR$ accepts $w$.
  \end{itemize}
\end{proof}

\subsection{Characterization of automata recognizing simple
  sets}\label{subsec:char:aut}
The first main theorem of this paper is now given.
\begin{theorem}\label{theo:aut-R}
  It is decidable in time $\bigO{nb}$ whether a minimal weak
  {}Büchi RNA with $n$ states recognizes a simple set.
\end{theorem}
In order to simplify the proof, this theorem is reduced to a simpler
case given in the following proposition.
\begin{proposition}\label{prop:aut-R}
  It is decidable in time $\bigO{nb}$ whether a minimal automaton with
  $n$ states recognizes a simple set different from $\emptyset$ and from
  $[0,\infty)$.
\end{proposition}
As in
Section \ref{subsec:char:bounded-aut}, a set of automata $\BAR$ is now
introduced.  Four lemmas are then proved, which corresponds to the
four properties of Proposition \ref{prop:method}.
\begin{definition}[$\BAR$]\index{A@$\BAR$}\label{def:BAR}
  Let $\BAR$ be the set of weak {}Büchi automata
  $\mathcal{A}=\autPar{Q}{\digitDotSet}{\delta}{\iniState}{F}$ such that:
 \begin{enumerate}
 \item\label{BAR-bar} The automaton $\mathcal{A}$ satisfies the properties of
   Definition \ref{def:BAF}. 
 \item\label{BAR-non-trivial} There exists an accepting and a
   rejecting strongly connected component, accessible from the initial
   state, belonging to $\fraStates$.
 \item\label{BAR-q0-0} $\del{q_{0}}{0}=q_{0}$.
 \item\label{BAR-empty-singleton} The set $\emptyStates$ contains
   exactly one recurrent state. Let $\emptyState$ denotes its only
   state.
 \item\label{BAR-inf-singleton} The set $\inftyStates$ contains at
   most one recurrent element. If $\inftyStates$ contains one recurrent
   element, let $\inftyState$ denote this only element.
 \item\label{BAR-q0-a}   
   $\del{q_{0}}{a}\ne\iniState$ for all $0<a<b$.
 \item\label{BAR-infty-or-empty} If $\inftyState$ exists, then
   $\del{q}{a}\ne\emptyState$ for all
   $q\in\natStates\setminus\set\emptyState$ and $a\in\digitSet$.
 \item\label{BAR-recurrent-nat}\label{BAR-last} Let $q$ be a natural
   recurrent state.
   The state $q$ is either $\emptyState$, $\inftyState$ or $q_{0}$.
 \end{enumerate}
\end{definition}
Note that the properties of Definition \ref{def:BAF} only consider the states
of $\fraStates$.  Therefore, there is no trouble to state that those
properties are satisfied on some subset of an automaton closed under
the function $\delta$.  Let us show that $\BAR$ admits the properties
of Proposition \ref{prop:method}. It is now proven that the set $\BAR$
satisfies Property \eqref{prop:method-algo} of Proposition \ref{prop:method}.
\begin{lemma}\label{lem:dans-BAR}
  It is decidable in time $\bigO{nb}$ whether a weak{} Büchi automaton
  $\mathcal{A}$ with $n$ states belong to $\BAR$.
\end{lemma}
\begin{proof}
  It suffices to check that all properties of Definition \ref{def:BAR} are
  testable in time $\bigO{nb}$. To check Property \eqref{BAR-bar}, it
  suffices to use the algorithm of Lemma \ref{lem:dans-baf}.

  The set of states accessible from $q_{0}$ is easily computed by a
  fixed-point algorithm in time $\bigO{nb}$. Using Tarjan's
  algorithm~\cite{Tarjan}, the set of strongly connected component are
  computable in time $\bigO{nb}$.  By Lemma \ref{lem:comp-set}, the sets
  $\emptyStates$ and $\inftyStates$ are computable in time
  $\bigO{nb}$. It easily follows that testing whether $\inftyState$
  exists is testable in time $\bigO{nb}$.  Since the set of recurrent
  states, the states $\emptyState$ and $\inftyState$, the set of
  states accessible from $q_{0}$, the sets $\emptyStates$,
  $\inftyStates$, $\natStates$ and $\fraStates$ are computed, it is
  trivial to test the seven last properties in time $\bigO{nb}$.
\end{proof}
It is now proven that the set $\BAR$ satisfies Property
\eqref{prop:method-AR} of Proposition \ref{prop:method}.
\begin{lemma}\label{lem:AR-in-BAR}
  Let $R\subsetneq[0,\infty)$ be a simple non-empty set.  The
  automaton $\AR$ belongs to $\BAR$.
\end{lemma}
\begin{proof}
  By Lemma \ref{lem:AR-in-BAF} and Lemma \ref{lem:quotient-BAF}, all
  automata $\mathcal{A}_{i}$ belongs to $\BAF$. Since Property \eqref{BAR-bar} only consider the states of $\fraStates$, that is the
  states of the form $(i,q)$ for $q\in Q_{i}$, then $\mathcal{A}_{R}$
  satisfies Property \eqref{BAR-bar}.

  Property \eqref{BAR-non-trivial} is now considered. Since $R$ is
  neither the empty set nor $[0,\infty)$, there exists a word
  belonging to $\digitSet^{*}\realDot\digitSet^{\omega}$ which is
  rejected by $\mathcal{A}_{R}$ and a word belonging to
  $\digitSet^{*}\realDot\digitSet^{\omega}$ which is accepted by
  $\mathcal{A}_{R}$. Therefore $\mathcal{A}_{R}$ satisfies Property \eqref{BAR-non-trivial}.
  
  Property \eqref{BAR-q0-0} is now considered.  Recall that the
  threshold of $R$, $t_{R}$ is positive.  Thus $b0+0<t_{R}$ and by
  construction of $\mathcal{A}_{R}$, $\del{\iniState}{0}=q_{b0+0}=q_{0}$,
  therefore $\mathcal{A}_{R}$ satisfies \ref{BAR-q0-0}.

  The automaton $\mathcal{A}_{R}$ satisfy Properties
  \ref{BAR-empty-singleton}, \ref{BAR-inf-singleton} and
  \ref{BAR-infty-or-empty} by construction.

  Property \eqref{BAR-q0-a} is now considered. Let $a>0$,
  $q\in\natStates$ and let us prove that $\del{q}{a}\ne q_{0}$. Two
  cases must be considered, depending on whether $t_{R}\ge a$ or
  whether $a<t_{R}$.  Let us first assume that $t_{R}\ge a$. Note that
  $b0+a\ge t_{R}$. By construction of $\mathcal{A}_{R}$, $\del{q_{0}}{a}$ is
  either $\inftyState[R]$ or is $\emptyState[R]$, which are distinct
  from $q_{\emptyState}$. Finally, let us assume that $a<t_{R}$.  By
  construction of $\mathcal{A}_{R}$ $\del{q_{0}}{a}=q_{b0+a}=q_{a}\ne q_{0}$,
  therefore $\mathcal{A}_{R}$ satisfy Property \eqref{BAR-q0-a}.

  Property \eqref{BAR-recurrent-nat} is now considered. By construction
  of $\mathcal{A}_{R}$, there are at most three recurrent states in
  $\natStates$: the initial state, the state
  $\emptyState[R]\in\emptyStates[\mathcal{A}_{R}]$, and potentially
  $\inftyState[R]\in\inftyStates[\mathcal{A}_R]$. Therefore $\mathcal{A}_{R}$
  satisfy Property \eqref{BAR-recurrent-nat}.
\end{proof}

It is now proven that the set $\BAR$ satisfies Property
\eqref{prop:method-min} of Proposition \ref{prop:method}.
\begin{lemma}\label{lem:quotient-BAR}
  The minimal quotient of automata of $\BAR$ belong to $\BAR$.
\end{lemma}
\begin{proof}
  Let $\mathcal{A}=\autPar{Q}{\digitDotSet}{\delta}{\iniState}{F}$ belonging
  to $\BAR$ and let
  $\mathcal{A}'=\autPar{Q'}{\digitDotSet}{\delta'}{q'_{0}}{F'}$ be a quotient
  of $\mathcal{A}$ by a morphism $\mu$. It is now shown that $\mathcal{A}'$ belongs
  to $\BAR$. Each of the \ref{BAR-last} propositions of
  Definition \ref{def:BAR} are considered separately.

  Property \eqref{BAR-bar} is first considered. By
  Lemma \ref{lem:quotient-BAF}, the set of automata satisfying Property \eqref{BAR-bar} is closed under quotient. Since $\mathcal{A}$ satisfies
  Property \eqref{BAR-bar}, and since $\mathcal{A}'$ is a quotient of $\mathcal{A}$,
  $\mathcal{A}'$ satisfies Property \eqref{BAR-bar}.

  Property \eqref{BAR-non-trivial} is now considered.  It is now proven
  that there exists an accepting recurrent state $q'$ accessible from
  the initial state of $\mathcal{A}'$. The case of a non-accepting
  state is similar. By Property \eqref{BAR-non-trivial}, there exists an
  accepting recurrent state $q$ accessible from $q_{0}$, in
  $\fraStates$. Since $q$ is accessible from $q_{0}$, there exists a
  finite word such that $\del{q_{0}}{w}=q$. Since $q$ is a recurrent,
  there exists a non-empty word $v$ such that $\del{q}{v}=q$. Since
  $q$ is recurrent and accepting, it does not belongs to
  $\emptyStates$. Since $q$ is fractional, since $\mathcal{A}$ accepts
  a subset of $\digitSet^{*}\realDot\digitSet^{\omega}$, and since
  $q\not\in\emptyStates$, then
  $w\in\digitSet^{*}\realDot\digitSet^{*}$.  Note that
  $\dell{\mu(q)}{v}=\mu(\del{q}{v})=\mu(q)$, therefore $\mu(q)$ is
  recurrent. By Lemma \ref{lem:morphism-sets}, $\mu(q)$ is
  fractional. Furthermore
  $\dell{q'_{0}}{w}=\dell{\mu(q_{0})}{w}=\mu(\del{q_{0}}{w})=\mu(q)$,
  therefore $q$ is accessible from $q'_{0}$.  Since $q$ is an
  accepting recurrent state of $\fraStates[\mathcal{A}']$, accessible
  from $q'_{0}$, $\mathcal{A}'$ satisfies Property \eqref{BAR-non-trivial}.

  Property \eqref{BAR-q0-0} is now considered.  Since
  $\dell{q'_{0}}{0}=\dell{\mu(q_{0})}{0}=
  \mu(\del{q_{0}}{0})=\mu(q_{0})=q'_{0}$,
  the automaton $\mathcal{A}'$ satisfies Property \eqref{BAR-q0-0}.

  Properties \ref{BAR-empty-singleton} and \ref{BAR-inf-singleton} are
  now considered. By Lemma \ref{lem:min-empty-infty}, since
  $\mathcal{A}'$ is minimal, there is at most one state in
  $\inftyStates[\mathcal{A}']$ and in
  $\emptyStates[\mathcal{A}']$. Furthermore
  $\del{q_{0}}{\realDot\realDot}$ belongs to
  $\emptyStates[\mathcal{A}']$, hence $\emptyStates[\mathcal{A}']$ is
  not empty. Hence $\mathcal{A}'$ satisfies Properties
  \ref{BAR-empty-singleton} and \ref{BAR-inf-singleton}.

  Property \eqref{BAR-q0-a} is now considered. Let $0<a<b$, it is now
  proven that $\dell{q'_{0}}{a}\ne q'_{0}$. By Property \eqref{BAR-q0-a}, $\del{q_{0}}{a}\ne q_{0}$.  By
  Lemma \ref{lem:run-recurrent}, there exists $i\in\mathbb N^{>0}$ such that
  $\del{q_{0}}{a^{i}}$ is recurrent. Let $q=\del{q_{0}}{a^{i}}$. By
  Property \eqref{BAR-recurrent-nat}, $q$ is either $q_{0}$,
  $\emptyState$ or $\inftyState$. The three cases are considered
  separately.
  \begin{itemize}
  \item It is first assumed that $q=q_{0}$, in this case, all of the
    states $\del{q_{0}}{a^{i}}$ are in the same strongly connected
    component. And since $\del{q_{0}}{a}\ne q_{0}$, Property \eqref{BAR-recurrent-nat} implies that $\del{q_{0}}{a}$ is either
    $\emptyState$ or $\inftyState$. The case where $\del{q_{0}}{a}$ is
    $\emptyState$ is now considered. The other case is similar. Since
    $\emptyState$ and $\iniState$ are in the same strongly connected
    component, $\iniState$ is accessible from $\emptyState$. Therefore
    $\mathcal{A}$ recognizes the empty language. Having $\mathcal{A}$ recognizing
    the empty language contradicts Property \eqref{BAR-non-trivial}.
  \item It is now assumed that $q = \emptyState$, the case
    $q=\inftyState$ is similar. By Property \eqref{BAR-empty-singleton},
    since $q=\emptyState$, $\mu(q)\in\emptyStates[\mathcal{A}']$.
    Furthermore,
    $\mu(q)=\mu(\del{q_{0}}{a^{i}})=\dell{\mu(q_{0})}{a^{i}}=\dell{q'_{0}}{a^{i}}$.
    Note that $\dell{q'_{0}}{a^{i}}=\inftyState[\mathcal{A}']\ne q'_{0}$.
    Since $\dell{q'_{0}}{a^{i}}\ne q'_{0}$ it follows that
    $\dell{q'_{0}}{a}\ne q'_{0}$. Therefore, $\mathcal{A}'$ satisfies
    Property \eqref{BAR-q0-a}.
  \end{itemize}

  Property \eqref{BAR-infty-or-empty} is now considered. Let us assume
  that $\inftyState[\mathcal{A}']$ exists.  Let $q'\in Q'$ and
  $a\in\digitSet$ such that
  $\dell{q'}{a}=\emptyState[\mathcal{A}']$. It must be proven that
  $q'\not\in\natStates[\mathcal{A}']\setminus\set{\emptyState[\mathcal{A}']}$. Since
  the automaton is minimal, the strongly connected component of
  $\inftyState[\mathcal{A}']$ is $\set{\inftyState[\mathcal{A}']}$.
  By Lemma \ref{lem:scc-morphism-codomain}, there exists a strongly
  connected component $C$ in $\mathcal{A}$ such that
  $\mu(C)=\set{\inftyState[\mathcal{A}']}$. Let $\inftyState$ be a
  state of $C$, since $\mathcal{A}_{\inftyState}$ and
  $\mathcal{A}'_{\mu(\inftyState)}=\mathcal{A}'_{\inftyState[\mathcal{A}']}$
  recognizes the same language, and since
  $\inftyState[\mathcal{A}']\in\inftyStates[\mathcal{A}']$ then
  ${\inftyState}\in \inftyStates$. Since $\inftyState$ belongs to a
  strongly connected component, it is recurrent. Since $\inftyState$
  is a recurrent state belonging to $\inftyState$, by Property \eqref{BAR-infty-or-empty}, $\del{q}{a}\ne\emptyState$ for all
  $q\in\natStates\setminus\set{\emptyState}$ and $a\in\digitSet$.  By
  definition of morphism, there exists a state $q\in Q$ such that
  $\mu(q)=q'$. By Lemma \ref{lem:morphism-sets}, since
  $q'\in\natStates[\mathcal{A}']$, $q\in\natStates$. Note that
  $\mu(\del{q}{a})=\dell{\mu(q)}{a}=\dell{q'}{a}=\emptyState[\mathcal{A}']\in\emptyStates[\mathcal{A}']$.
  By Lemma \ref{lem:morphism-sets}, since
  $\mu(\del{q}{a})\in\emptyStates[\mathcal{A}']$,
  $\del{q}{a}\in\emptyStates$. Since $\del{q}{a}\in\emptyStates$, by
  Property \eqref{BAR-empty-singleton}, $\del{q}{a}=\emptyState$. It
  implies that $q\not\in\natStates\setminus\set{\emptyState}$. By
  Lemma \ref{lem:morphism-sets}, it implies that
  $q'\not\in\natStates[\mathcal{A}']\setminus\set{\emptyState[\mathcal{A}']}$. Therefore,
  $\mathcal{A}'$ satisfies Property \eqref{BAR-infty-or-empty}.
  
  Finally, Property \eqref{BAR-recurrent-nat} is now considered.  Let
  $q'$ be a natural recurrent state of $\mathcal{A}'$. It must be proven that
  $q'$ is either $q_{0}$, $\emptyState[\mathcal{A}']$ or
  $\inftyState[\mathcal{A}']$.  The state $q'$ is natural, by
  Lemma \ref{lem:morphism-sets}, its antecedents by $\mu$ are natural.
  The state $q'$ is recurrent hence, by
  Lemma \ref{lem:scc-morphism-codomain}, it admits a recurrent
  antecedent $q$. By Property \eqref{BAR-recurrent-nat} applied to
  $\mathcal{A}$, either $q=q_{0}$, $q=\emptyState$ or $q=\inftyState$. The
  three cases are considered separately.
  \begin{itemize}
  \item The case $q=q_{0}$ is first considered, in this case, clearly,
    $q'=q'_{0}$.
  \item The case $q=\emptyState$ is now considered. As proven above,
    it implies that $\mu(q)=\emptyState[\mathcal{A}']$.
  \item The case where $q=\inftyState$ is similar to the preceding
    case.
  \end{itemize}
  Therefore, $\mathcal{A}'$ satisfies Property \eqref{BAR-recurrent-nat}.
\end{proof}
It is now proven that the set $\BAR$ satisfies Property
\eqref{prop:method-L} of Proposition \ref{prop:method}.
\begin{lemma}\label{BAR->pour}
  The automata of $\BAR$ recognize simple sets.
\end{lemma}
In order to prove this lemma, another lemma is first introduced. It
implies that the set $R$ recognized by an automaton $\mathcal{A}\in\BAR$ with
$n$ states is such that $[0,b^{n})$ is either a subset of $R$ or is
disjoint from $R$.
\begin{lemma}\label{lem:cycle-N}
  Let $\mathcal{A}\in\BAR$ be an automaton with $n$ states recognizing a set
  $R$.  If $\mathcal{A}$ contains a state $\inftyState$, as in
  Definition \ref{def:BAR}, then $(b^{n-1},\infty)\subseteq R$, otherwise
  $(b^{n-1},\infty)\cap R=\emptyset$.
\end{lemma}
\begin{proof}
  Two cases must be considered, depending on whether the state
  $\inftyState$ exists.  Let us assume that the state $\inftyState$
  exists, the other case is similar.  Let $x>b^{n-1}$ and let
  $0^{c}\natPart{w}\realDot \fraPart{w}$, be one of its encoding in base
  $b$, with $c\in\mathbb N$ and $\natPart{w}[0]\ne0$. Let us prove that $\mathcal A$
  accepts $0^{c}\natPart{w}\realDot \fraPart{w}$

  Note that since $x>b^{n-1}$, it implies that the length of
  $\natPart{w}$ is at least $n$. For $i\le\length{\natPart{w}}$, let
  $q_{i}^{\nat}=\del{\iniState}{0^{c}\prefix{\natPart{w}}{i}}$.  By Lemma
  \ref{lem:run-recurrent}, there exists $0\le i'<i\le n$ such that
  $q^{\nat}_{i}$ is recurrent. By Property \eqref{BAR-recurrent-nat} of
  Definition \ref{def:BAR}, the only natural recurrent states of
  $\mathcal{A}$ are $\inftyState$, $\emptyState$ and $q_{0}$. By
  Properties \ref{BAR-q0-a} and \ref{BAR-infty-or-empty},
  $q^{\nat}_{i}\ne q_{0}$.  Since $\inftyState$ exists, by Property \eqref{BAR-infty-or-empty} then ${q^{\nat}_{i}}\ne\emptyState$.  Since
  $q^{\nat}_{i}$ is either $\inftyState$, $\emptyState$, or $q_{0}$,
  since ${q^{\nat}_{i}}\ne\emptyState$ and since
  $q^{\nat}_{i}\ne q_{0}$ $q^{\nat} _{i}=q_{\inftyState}$. It follows
  that $\del{q^{\nat}_{i}}{\natPart{w}}= \inftyState$, and then, for all
  $j\in\mathbb N$,
  $\del{q^{\nat}}{0^{c}\natPart{w}\realDot\prefix{\fraPart{w}}{j}}=
  \zuState$.
  Therefore $\mathcal{A}$ accepts $0^{c}\natPart{w}\realDot \fraPart{w}$.
\end{proof}
Example \ref{BAR->pour} is now proven
\begin{proof}[Proof of Example \ref{BAR->pour}]
  Let $\mathcal{A}\in \BAR$ with $n$ states and let
  $R=\wordToReal{\mathcal{A}}$. Let us prove that $R$ is simple.  By
  Lemma \ref{lem:cycle-N}, it suffices to prove that $R\cap[0,b^{n-1})$ is
  simple. In order to do this, it suffices to prove that $R_{i}$ is
  simple for all $i\in[b^{n-1}-1]$.

  By Property \eqref{BAR-bar} of Definition \ref{def:BAR},
  $\mathcal{A}_{\del{\iniState}{w\realDot}}\in\BAF$ for all
  $w\in\digitSet^{*}$.  Note that the difference between $R_{i}$ and
  $R'_{i}=\bigcup_{j\in\mathbb N}\wordToReal{\mathcal{A}_{\del{\iniState}{0^{j}w\realDot}}}$
  is a subset of $\set{0,1}$. Therefore, it suffices to prove that
  $R'_{i}$ is simple. Note that, since $\del{q_{0}}{0}=q_{0}$, all
  $\mathcal{A}_{\del{\iniState}{0^{j}w\realDot}}$ are equals. Therefore the
  infinite union $R'_{i}$ of simple sets is a simple set. That is,
  $R'_{i}$ is a simple set.
\end{proof}
Proposition \ref{prop:method} is now proven. Its proof is similar to the proof
of Theorem \ref{theo:aut-0-1}.
\begin{proof}[Proof of Proposition \ref{prop:method}]
  It suffices to use Proposition \ref{prop:method} with $\mathbb A$ being the set
  of automata $\BAR$, $\mathbb A'$ being the set of RNA, and Lemmas
  \ref{lem:dans-BAR}, \ref{lem:AR-in-BAR}, \ref{lem:quotient-BAR} and
  \ref{BAR->pour}.
\end{proof}
Theorem \ref{theo:aut-R} is now proven.
\begin{proof}[Proof of Theorem \ref{theo:aut-R}]
  The algorithm to decide whether a minimal weak
  {}Büchi RNA $\mathcal{A}$ recognizes a simple set
  is now given.  The algorithm first checks whether Property \eqref{BAR-non-trivial} of Definition \ref{def:BAR} holds. If it does not
  hold, then the automaton recognizes the empty language or
  $\digitSet^{*}\realDot\digitSet^{\omega}$, in which cases the
  algorithm accepts.  If Property \eqref{BAR-non-trivial} holds, then
  the automaton recognizes a non empty strict subset of
  $\digitSet^{*}\realDot\digitSet^{\omega}$ and therefore the problem
  is reduced to the problem considered in
  Proposition \ref{proposition:bounded-construct}. It thus suffices to apply
  the algorithm of Proposition \ref{proposition:bounded-construct} to $\mathcal{A}$
  and to return the result of this algorithm.
\end{proof}

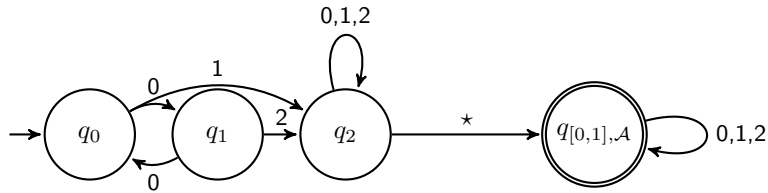
\begin{figure}[h]
  \centering
    \begin{tikzpicture}[->, >=stealth', shorten >=1pt, auto, node
      distance=1.7cm, thick, main node/.style={circle, fill=!20, draw,
        font=\sffamily\Large\bfseries, inner sep=0.1pt, minimum
        size=1cm}] \tikzset{every state/.style={minimum size=1.2cm}}
      \node[state, initial, initial text={}] (even) {$\iniState$};
      \node[state, right of=even] (odd) {$q_{1}$};
      \node[state, right of=odd] (acc) {$q_{2}$};
      \node[state, right=2cm of acc, accepting] (01) {$\zuState$};

      \path[every node/.style={font=\sffamily\small}]
      (even)  edge [bend left] node {0} (odd)
      (odd)  edge [bend left] node {0} (even)
      (even) edge [bend left] node {1} (acc)
      (odd) edge  node {2} (acc)
      (acc) edge [loop above] node {0,1,2} (acc)
      (acc) edge node {$\realDot$} (01)
      (01) edge [loop right] node {0,1,2} (01)
           ;
    \end{tikzpicture}
    \caption{An automaton which recognizes $[1,\infty)$ and is refused by
    the algorithm of Theorem \ref{theo:aut-R}.}
    \label{fig:false-neg}
  \end{figure}

The algorithm of Theorem \ref{theo:aut-R} takes as input a RNA and runs in
time $\bigO{nb}$. It should be noted that it is not known whether it
is decidable in time $\bigO{nb}$ whether an automaton is a RNA.
However, as for the algorithm of Theorem \ref{theo:aut-0-1}, if the
algorithm of Theorem \ref{theo:aut-R} is applied to a weak
{}Büchi automaton which is not a real
automaton, the algorithm returns no false positive.  An example of
false negative is now given.  Let $L$ be the language described by the
regular expression:
\begin{equation*}
  \left(00\right)^{*}\left(01+2\right)\digitSet[3]^{*}\realDot
  \digitSet[3]^{\omega}.
\end{equation*}
This language is recognized by the automaton of
Figure \ref{fig:false-neg}.  Note that $\wordToReal{L}=[1,\infty)$, which
is a simple set. However, since Property \eqref{BAR-q0-0} is not
satisfied by the automaton of Figure \ref{fig:false-neg}, the algorithm
of Theorem \ref{theo:aut-R} does not accept this automaton.

\section{From automata to simple set}\label{sec:aut->set}
In this section, it is explained, given a weak
{}Büchi automaton recognizing a simple set $R$,
how to compute a first-order formula which defines $R$. The exact
theorem is now stated.

\begin{theorem}\label{theo:bounded-construct}
  Let $\mathcal{A}$ be a be a minimal RNA with $n$ states, over the alphabet
  $\digitSet$, which recognizes a simple set. There exists two
  formulas which define $\wordToReal{\mathcal{A}}$:
  \begin{itemize}
  \item a $\ef{\mathbb R;+,<,1}$-formula computable in time
    $\bigO{n^{2}b\log(nb)}$ and
  \item a $\sigF{2}{\mathbb R;+,<,1}$-formula computable in time
    $\bigO{nb\log(nb)}$.
  \end{itemize}
\end{theorem}

In order to prove this theorem, a more general proposition is
introduced. This proposition shows that the condition that $\mathcal{A}$ is a
RNA can be replaced by the condition $\mathcal{A}\in\BAR$.
\begin{proposition}\label{proposition:bounded-construct}
  Let $\mathcal{A}\in\BAF$ be a minimal automaton.  There exists two formulas
  which define $\wordToReal{\mathcal{A}}$:
  \begin{itemize}
  \item a $\ef{\mathbb R;+,<,1}$-formula computable in time
    $\bigO{n^{2}b\log(nb)}$ and
  \item a $\sigF{2}{\mathbb R;+,<,1}$-formula computable in time
    $\bigO{nb\log(nb)}$.
  \end{itemize}
\end{proposition}

In order to prove this proposition, a technical lemma is first
introduced.  
\begin{lemma}\label{lem:C}
  Let $A\in\BAR$ be minimal with $n$ states, $\natPart{w}\in\digitSet^{*}$,
  $\fraPart{w}\in\digitSet^{\omega}$ and $\mathcal{Q}\subseteq Q$ containing
  exactly one state of each strongly connected component. Then, there
  exists $s\in[n]$ such that
  $\del{q}{\natPart{w}\realDot(\prefix{\fraPart{w}}s)}\in \mathcal{Q}$.
\end{lemma}
\begin{proof}
  Let
  $q_{i}^{\fract}=\del{\iniState}{\natPart{w}\realDot(\prefix{\fraPart{w}}i)}$
  for any $i\in\mathbb N$. By Lemma \ref{lem:run-recurrent}, there exists
  $0\le i<i'\le n$ such that $q_{i}^{\fract}$ is recurrent. Let $C$ be
  the strongly connected component of $q_{i}^{\fract}$. By Property \eqref{BAR-recurrent-nat} of Definition \ref{def:BAR}, there are three kinds
  of strongly connected components in the fractional part of a minimal
  automaton of $\BAR$: the singleton $\set{\emptyState}$, the
  singleton $\set{\zuState}$, and the cycles. Three cases must be
  considered depending on the kind of strongly connected components
  that is $C$. In the two first cases, if $q^{\fract}_{i}$ is
  $\emptyState$ or $\zuState$ then $q^{\fract}_{i}\in\mathcal{Q}$ and it
  suffices to take $s=i$.  Otherwise, if $C$ is a cycle, then
  $\set{q^{\fract}_{j}\mid i\le j<i'}=C$, therefore there exists an
  integer $i\le s<i'$ such that $q^{\fract}_{s}\in C$.
\end{proof}

Proposition \ref{proposition:bounded-construct} can now be proven.
\begin{proof}
  Let $R=\wordToReal{\mathcal{A}}$. Since the automaton $\mathcal{A}$
  belongs to $\BAR$, the notations of Definition \ref{def:BAR} can now be
  used, and therefore the notations of Definition \ref{def:BAF} can also
  be used. Recall that all strongly connected components are cycles,
  apart from $\set\iniState$, $\set\inftyState$, $\set\zuState$ and
  $\set\emptyState$. Furthermore, for each $q\in\fraStates$ in a cycle
  $C$, the digit $\suc{q}$ is the only one such that
  $\del{q}{\suc{q}}\in C$.  In this proof, it is assumed that each
  state has an integer index between $0$ and $n-1$. For $q$ a state
  with index $i$ and $p$ a variable, the atomic first-order formula
  $p\doteq q$ is an abbreviation for $p\doteq i$.  Let
  $\mathcal{Q}\subseteq Q$ be a set as in Lemma \ref{lem:C}.  Note
  that the states $\emptyState$, $\zuState$ and $\inftyState$ all
  belong to $\mathcal{Q}$ if they exists.

  It is first explained how to compute an existential formula
  $\phi(x)$ which defines $R$ in time
  $\bigO{n^{2}\log(n)b\log(b)}$. As seen in Lemma \ref{lem:cycle-N},
  the formula which defines $R$ is either $\phi(x)\lor x\ge b^{n-1}$
  if $\inftyState$ exists, or $\phi(x)$ otherwise.  At the end of the
  proof, it is explained how to decrease the time by adding universal
  quantifiers.

  For any real $0\le x<b^{n-1}$, let $\natPart{x}\in\mathbb N$ and
  $x^{\fract}\in[0,1]$ be numbers such that
  $x=\natPart{x}+\fraPart{x}$. If $x\in\mathbb N^{>0}$, then the pair
  $\left(\natPart{x},\fraPart{x}\right)$ is either
  $\left(\floor{\natPart{x}},x-\floor{\natPart{x}}\right)$ or $\left(x-1,1\right)$.
  Otherwise, the pair $\left(\natPart{x},\fraPart{x}\right)$ is
  $\left(\floor{\natPart{x}},x-\floor{\natPart{x}}\right)$. Let
  $w_{\natPart{x}}\in\digitSet^n$ be an encoding in base $b$ of $\natPart{x}$ of
  length $n$. Since $\natPart{x}<b^{n-1}$, such an encoding exists. Let
  $w_{\fraPart{x}}\in\digitSet^{\omega}$ be an encoding in base $b$ of
  $\fraPart{x}$.  Let
  $q^{\nat}_{x,i}=\del{\iniState}{\prefix{w_{\natPart{x}}}{i}}$ for all
  $i\in[n]$ and
  $q^{\fract}_{x,i}=\del{\iniState}{w_{\natPart{x}}\realDot(\prefix{w_{\fraPart{x}}}
    i)}$ for all $i\in\mathbb N$.

  \paragraph{}
  The formula $\phi\left(x\right)$ which defines $\wordToReal{\AR}$ is
  defined as the conjunction of two subformulas.  Intuitively, the
  first formula, $\phi_{\nat}\left(p^{\nat},\natPart{x}\right)$ considers
  the run on $\natPart{w}$ and the second formula,
  $\fraPart{\phi}\left(p^{\fract},\fraPart{x}\right)$, considers the run
  on $\fraPart{w}$.

  Let us assume that there exists a formula
  $\phi_{\nat}\left(p^{\nat},\natPart{x}\right)$, of size
  $\bigO{n^{2}b\log\left(nb\right)}$, such that, if
  $\natPart{x_{n}}<b^{n-1}$, the formula holds if and only if $p^{\nat}$ is the
  index of $\fraPart{w_{x}}$ and if $\fraPart{x}\in\mathbb N$.  Let us assume
  that there exists a formula
  $\fraPart{\phi}\left(p^{\fract},\fraPart{x}\right)$, of size
  $\bigO{n^{2}b\log(nb)}$, which accepts $\fraPart{x}\in[0,1]$ if and
  only if $\mathcal{A}_{p^{\nat}}$ accepts an encoding of $\fraPart{x}$.  Then it
  suffices to take $\phi(x)$ to be the formula:
  \begin{equation*}
    \arraycolsep=0.5pt
    \begin{array}{llll}
      \phi(x)=\exists    \natPart{x},\fraPart{x},p^{\nat},p^{\fract}.
      &      x\doteq \natPart{x}+\fraPart{x}
      \land \natPart{x}<b^{n-1}\land
              \fraPart{x}\in[0,1]
      \\
      &\land\bigvee_{q\in \natStates}\left(p^{\nat}\doteq{}q\land p^{\fract}\doteq{}\del{q}{\realDot}\right)
      \\
      &\land
              \phi_{\nat}\left(p^{\nat},\natPart{x}\right)
              \land
              \fraPart{\phi}\left(p^{\fract},\fraPart{x}\right).
    \end{array}
  \end{equation*}
  \paragraph{}
  The formula $\phi_{\nat}\left(p^{\nat},\natPart{x}\right)$ is now
  defined.  A sequence $\left(\natPart{x_{i}}\right)_{i\in[n]}$ of
  variables is existentially quantified in this formula. The variable
  $\natPart{x_{i}}$ is intended to be interpreted by the value
  $\floor{\frac{\natPart{x}}{b^{n-i}}}$, its encoding in base $b$ is
  $\prefix{w_{\natPart{x}}}{i}$.  A sequence
  $\left(p_{i}^{\nat}\right)_{i\in[n]}$ of variables is existentially
  quantified. They are used to encode the $n$ steps of the run of
  $\mathcal{A}$ on $\natPart{w}$. More precisely, the variable
  $p^{\nat}_{i}$ is meant to be interpreted by the index of
  $q_{i}^{\nat}$.  Note that if $\natPart{x_{0}}=0$, since
  $\natPart{x_{i+1}}=b \natPart{x_{i}}+w_x[i]$ for $i\in[n-1]$, an easy
  induction shows that $\natPart{x_{i}}\in\mathbb N$ for all $0\le i\le n$.

  Let
  us assume that there exists a $\qf{\mathbb R;+,<,1}$-formula
  $\psi^{\nat}\left(p^{\nat}_{i},\natPart{x_{i}},p^{\nat}_{i+1},x_{i+1}^{\nat}\right)$,
  of size $\bigO{nb\log(nb)}$, which asserts that if $p_{i}$ is the
  index of a state $q$, and if $\natPart{x_{i+1}}= b\natPart{x_{i}}+a$ for
  some $a\in\digitSet$, then $p_{i+1}$ is the index of $\del{q}{a}$.
  The formula $\phi_{\nat}\left(p^{\nat},\natPart{x}\right)$ can then be
  taken to be the $\ef{\mathbb R;+,<,1}$-formula of length
  $\bigO{n^{2}b\log\left(nb\right)}$:
  \begin{equation*}
    \arraycolsep=0.5pt
    \begin{array}{ll}
    \phi_{\nat}\left(p^{\nat},\natPart{x}\right)
    =
    \exists \left(p_{i}^{\nat},\natPart{x_{i}}\right)_{0\le{i}\le{n}}.&
    \natPart{x_{0}}\doteq0\land p_{0}^{\nat}\doteq \iniState
    \\&\land
    \bigwedge_{i=0}^{n-1}\natPart{\psi}\left(p^{\nat}_{i},\natPart{x_{i}},p^{\nat}_{i+1},\natPart{x_{i+1}}\right)
    \\&\land \natPart{x_{n}}\doteq \natPart{x}\land p_{n}^{\nat}\doteq p^{\nat}.
    \end{array}
  \end{equation*}

  Formally, the formula
  $\natPart{\psi}\left(p^{\nat}_{i},\natPart{x_{i}},p^{\nat}_{i+1},\natPart{x_{i+1}}\right)$
  can be taken to be:
  \begin{equation*}
    \natPart{\psi}\left(
      \begin{array}{l}
        p^{\nat}_{i},\natPart{x_{i}},\\p^{\nat}_{i+1},\natPart{x_{i+1}}
     \end{array}
    \right) =
    \bigwedge_{q\in{\natStates}}\bigwedge_{a\in\digitSet}
    \left[
      \begin{array}{l}
      \left(
        \natPart{x_{i+1}}\doteq b\natPart{x_{i}}+a\land p^{\nat}_{i}\doteq
        q\right)\implies{}\\{p^{\nat}_{i+1}\doteq \del{q}{a}}
      \end{array}
    \right]
    .
  \end{equation*}
  No notations introduced during the construction of the formula
  $\phi_{\nat}\left(p^{\nat},\natPart{x}\right)$, are used in the
  remaining of the proof.
  \paragraph{}
  The formula $\fraPart{\phi}\left(p^{\fract},\fraPart{x}\right)$ is now
  defined.  Let $s$ be the smallest integer such that
  $q^{\fract}_{x,s}\in \mathcal{Q}$, by Lemma \ref{lem:C}, such an integer
  exists.  The formula
  $\fraPart{\phi}\left(p^{\fract},\fraPart{x}\right)$ is the conjunction
  of two subformulas. The first formula,
  $\fraPart{\phi_{1}}\left(p,y,p^{\fract},\fraPart{x}\right)$, considers
  the run of $\mathcal{A}$ on the first $s$ letters of $\fraPart{w_{x}}$.  The
  second formula $\fraPart{\phi_{2}}(p,y)$ considers the end of the
  run, on $\suffix{w_x}{s}$, beginning at the state
  $q_{x,s}\in\mathcal{Q}$.

  Two variables $p$ and $y$ are existentially quantified. They are
  meant to be interpreted by $p^{\fract}_{s}$ and
  $\suffix{\fraPart{w_{x}}}{s}$ respectively.  Assume that there is a
  $\ef{\mathbb R;+,<,1}$-formula,
  $\fraPart{\phi_{1}}\left(p,y,p^{\fract},\fraPart{x}\right)$, of length
  $\bigO{n^{2}b\log(b)}$, which states that, given $p^{\fract}$,
  $\fraPart{x}$'s, the variables $p$, $y$ are interpreted as stated
  above. Let us also assume that there exists a
  $\qf{\mathbb R;+,<,1}$-formula $\fraPart{\phi_{2}}(p,y)$, of length
  $\bigO{n^{2}\log(b)}$, which states that $\mathcal{A}_{q}$ accepts
  an encoding of $y$, where $p$ is the index of the state $q$.  Then
  the formula $\fraPart{\phi}$ can be taken to be the
  $\ef{\mathbb R;+,<,1}$-formula of length $\bigO{n^{2}b\log(nb)}$:
  \begin{eqnarray*}
    \fraPart{\phi}\left(p^{\fract},\fraPart{x}\right)
    =
    \exists
    p,y.
    \fraPart{\phi_{1}}\left(p,y,p^{\fract},\fraPart{x}\right)
    \land
    \fraPart{\phi_{2}}(p,y).
  \end{eqnarray*}

  \paragraph{}

  The formula,
  $\fraPart{\phi_{1}}\left(p,y,p^{\fract},\fraPart{x}\right)$ is now
  constructed.  As in the construction of
  $\fraPart{\phi}\left(p^{\fract},\fraPart{x}\right)$, two sequences of
  variables are existentially quantified to encode some suffix of
  $\fraPart{w}$ and to encode a part of the run of the automaton on
  $w$.  The sequence $\left(\fraPart{x_{i}}\right)_{i\in[n]}$ is
  existentially quantified. The variable $\fraPart{x_{i}}$ is meant to
  be interpreted by $\wordToReal{\suffix{w_{\fraPart{x}}}{i}}$. It
  represents the real that must be read at the $i$-th step of the run
  after the $\realDot$. It follows that $\fraPart{x_{0}}=\fraPart{x}$,
  $\fraPart{x_{s}}=y$ and that $\fraPart{x_{i}}$ is equal to
  $bx_{i-1}-w_{\fraPart{x}}[i]$.  The sequence
  $\left(p_{i}^{\fract}\right)_{i\in[n]}$ of variables is
  existentially quantified. It is used to encode the first $(n+1)$ steps
  of the run after the $\realDot$ part of the run. More precisely, the
  variable $p^{\fract}_{i}$ is meant to be interpreted by the indexes
  of $q_{i}^{\fract}$.  A third sequence of variables,
  $\left(s_{i}\right)_{i\in[n]}$, is existentially quantified. The
  variable $s_{i}$ is meant to be interpreted by $0$ if $i< s$ and $1$
  otherwise. Note that $s_{0}=0$ as $s\ge 0$.  Those variables allows
  to know the value of $s$.

  Let
  $\psi^{\fract}\left(p,y,p^{\fract}_{i},\fraPart{x_{i}},s_{i},p^{\fract}_{i+1},\fraPart{x_{i+1}},s_{i+1}\right)$
  be $\qf{\mathbb R;+,<,1}$-formula of length $\bigO{nb\log(nb)}$ which
  states that, given $p^{\fract}_{i},\fraPart{x_{i}},s_{i}$, the
  variables $p^{\fract}_{i+1}$, $\fraPart{x_{i+1}}$ and $s_{i+1}$ are
  correctly interpreted, and furthermore if $i=s$ -- that is if
  $p^{\fract}_{i}\in\mathcal{Q}$ and for all $j<i$,
  $p^{\fract}_{j}\not\in\mathcal{Q}$ -- then the variables $p$ and $y$
  are correctly interpreted.  The formula
  $\fraPart{\phi_{1}}\left(p,y,p^{\fract},\fraPart{x}\right)$ can then
  be expressed as the $\ef{\mathbb R;+,<,1}$-formula of length
  $\bigO{n^{2}b\log(nb)}$:
  \begin{equation*}
    \arraycolsep=0.5pt
    \fraPart{\phi_{1}}\left(\begin{array}{l}p,y,\\p^{\fract},\fraPart{x}\end{array}\right)
    =
    \begin{array}{ll}
      \exists(p^{\fract}_{i},\fraPart{x_{i}},s_{i})_{i\in[n]}.
      &p^{\fract}\doteq p^{\fract}_{0}\land \fraPart{x}\doteq \fraPart{x_{0}}\land
      \\&\bigwedge_{i=0}^{n-1}\psi^{\fract}\left(
          \begin{array}{l}
            p,y,p^{\fract}_{i},\fraPart{x_{i}},\\s_{i},p^{\fract}_{i+1},\fraPart{x_{i+1}},s_{i+1}
          \end{array}
      \right).
    \end{array}
  \end{equation*}

  The formula
  $\fraPart{\psi}\left(p,y,p^{\fract}_{i},\fraPart{x_{i}},s_{i},p^{\fract}_{i+1},\fraPart{x_{i+1}},s_{i+1}\right)$
  is now given. It is the $\qf{\mathbb R;+,<,1}$-formula of length
  $\bigO{nb\log(nb)}$:
  \begin{equation*}
    \arraycolsep=0.5pt
    \fraPart{\psi}\left(\begin{array}{llll}p^{\fract}_{i},&\fraPart{x_{i}},&s_{i},&p\\p^{\fract}_{i+1},&\fraPart{x_{i+1}},&s_{i+1}&y\end{array}\right)=
    \begin{array}{l}
      \left\{\left(s_{i+1}\doteq1\right)\iff{\left(s_{i}\doteq1\lor\bigvee_{q\in\mathcal{Q}}p^{\fract}_{i}\doteq{q}\right)}\right\}
      \land\\
      \left\{\bigvee_{q\in{Q}}\bigvee_{a=0}^{b-1} \left[
      \begin{array}{rcl c rcll}
        p^{\fract}_{i}&\doteq{}&q&\land&\fraPart{x_{i}}&\in&\left[\frac{a}{b},\frac{a+1}{b}\right]&
                                                                                                           \land\\
        p^{\fract}_{i+1}&\doteq{}&\delta(q,a)&\land& \fraPart{x_{i+1}}&\doteq{}&b\left(\fraPart{x_{i}}-\frac{a}{b}\right)&
      \end{array}
                                                                                                                      \right]\right\}\land\\
      \left\{\left[ s_{i+1}\doteq 1\land s_{i}\doteq0\right]\implies
      {\left[p\doteq p^{\fract}_{i}\land y\doteq \fraPart{x_{i}}\right]}\right\}.
    \end{array}
  \end{equation*}

  \paragraph{}
  The formula $\fraPart{\phi_{2}}(t,y)$ is now constructed. It is a
  disjunction, which states that there exists $q\in C$, such that
  $q_{x,s}=q$, and such that $\mathcal{A}_{q_{x,s}}$ accepts an encoding of
  $y=\fraPart{x_{s}}$. By definition of $s$, $q_{x,s}\in\mathcal{Q}$.
  Let us assume that, for each $q\in\mathcal{Q}$ in a strongly connected
  component $C$, there exists a $\qf{\mathbb R;+,<,1}$-formula $\xi_{q}(y)$
  of length $\bigO{\log(b)\card C}$, where $\card C$ is the cardinal
  of $C$, which states that $\mathcal{A}_{q}$ accepts an encoding of
  $y$. Then, $\fraPart{\phi_{2}}(p,y)$ can be taken as the
  $\qf{\mathbb R;+,<,1}$-formula of length $\bigO{n\log(nb)}$:
  \begin{equation*}
    \fraPart{\phi_{2}}(p,y)=\bigvee_{q\in\mathcal{Q}}p=q\land \xi_{q}(y).
  \end{equation*}

  Let us now construct the formula $\xi_{q}(y)$. Trivially,
  $\xi_{\zuState}(y)$ can be taken to be $\true$ and
  $\xi_{\emptyState}(y)$ can be taken to be $\false$. They are
  constant size formula.  It remains to construct the formulas
  $\xi_{q}(y)$ for $q$ in a cycle $C$. Let $s_{q}$ be the value
  $\suc{q}$ defined as in the proof of Lemma \ref{lem:BAF->pour}, for
  the automaton $\mathcal{A}_{q}$.  As shown in the proof of Lemma
  \ref{lem:BAF->pour}, the length of $\suc{q}$ is
  $\bigO{\log(b)\card C}$ and the automaton $\mathcal{A}_{q}$
  recognizes a set, which is a union of $[0,\suc{q})$,
  $\set{\suc{q}}$, and $(\suc{q},1]$, and which is defined by a
  formula $\xi_{q}$ of length $\bigO{\log(b)\card C}$.
  \paragraph{}
  It is now explained how to transform the $\ef{\mathbb R;+,<,1}$-formula
  $\phi(x)$ of length $\bigO{n^{2}b\log\left(nb\right)}$ into an
  equivalent $\sigF{2}{\mathbb R;+,<,1}$-formula of length
  $\bigO{nb\log(nb)}$. Let us assume that there exists
  $\phi^{\prime[0,1]}_{1}\left(p,y,p^{\fract},\fraPart{x}\right)$ and
  $\phi^{\prime\mathbb N}\left(p^{\nat},\natPart{x}\right)$, two
  $\sigF{2}{\mathbb R;+,<,1}$-formulas of length $\bigO{nb\log(nb)}$,
  equivalent to
  $\fraPart{\phi_{1}}\left(p,y,p^{\fract},\fraPart{x}\right)$ and to
  $\phi_{\nat}\left(p^{\nat},\natPart{x}\right)$ respectively.  It thus
  suffices to replace the two formulas
  $\fraPart{\phi_{1}}\left(p,y,p^{\fract},\fraPart{x}\right)$ and
  $\phi_{\nat}\left(p^{\nat},\natPart{x}\right)$ in $\phi(x)$ by their
  equivalent smaller formulas.

  In order to construct a $\sigF{2}{\mathbb R;+,<,1}$-formula of length
  $\bigO{nb\log(nb)}$ equivalent to
  $\fraPart{\phi_{1}}\left(p,y,p^{\fract},\fraPart{x}\right)$ or to
  $\phi_{\nat}\left(p^{\nat},\natPart{x}\right)$, it suffices to replace
  their last conjunctions by universal quantifications.  The formula
  $\fraPart{\phi_{1}}\left(p,y,p^{\fract},\fraPart{x}\right)$ is
  equivalent to the following $\sigF{2}{\mathbb R;+,<,1}$-formula of length
  $\bigO{nb\log(nb)}$:
  \begin{equation*}
    \fraPart{\phi_{1}}\left(
      \begin{array}{l}
        p,y,\\p^{\fract},\fraPart{x}
      \end{array}
    \right)=
    \begin{array}{ll}
      \exists
      (p^{\fract}_{i},\fraPart{x_{i}},s_{i})_{i\in[n]}. p^{\fract}\doteq p^{\fract}_{0}\land \fraPart{x}\doteq \fraPart{x_{0}}\land\\
      \forall    \arraycolsep=0.5pt
      \begin{array}{lll}
        \rho,&\xi,&\gamma,\\\rho',&\xi',&\gamma'
      \end{array}.
                                          \left\{
                                          \begin{array}{l}
                                            \left[\bigvee_{i=0}^{n-1}\left(
                                            \arraycolsep=0.5pt
                                            \begin{array}{llllll}
                                              \rho=p^{\fract}_{i}&\land& \xi=\fraPart{x_{i}}&\land&\gamma=\suc{i}&\land\\
                                              \rho'=p^{\fract}_{i+1}&\land& \xi'=\fraPart{x_{i+1}}&\land& \gamma'=\suc{i+1}&
                                            \end{array}
                                                                                                                            \right)
                                                                                                                            \right]
                                            \\\implies{} 
                                            \fraPart{\psi}\left(
                                            \arraycolsep=0.5pt
                                            \begin{array}{llll}
                                              \rho,&\xi,&\gamma,&t,\\\rho',&\xi',&\gamma',&y
                                            \end{array}
                                                                                            \right)
                                          \end{array}
                                                                                            \right\}.
    \end{array}
  \end{equation*}
  Similarly, the formula
  $\phi_{\nat}\left(p^{\nat}_{0},\natPart{x_{0}}\right)$ is equivalent to
  the $\sigF{2}{\mathbb R;+,<,0}$-formula of length $\bigO{nb\log(nb)}$:
  \begin{equation*}
    \phi_{\nat}\left(p^{\nat},\natPart{x}\right)=
    \exists \left(p^{\nat},\natPart{x_{i}}\right)_{0\le{i}\le{n}}.
    \begin{array}{ll}
      \natPart{x_{0}}\doteq0\land p_{0}^{\nat}\doteq \iniState\\
      \land \natPart{x_{n}}\doteq \natPart{x}\land p_{n}^{\nat}\doteq p^{\nat}
      \land\\
      \forall
      \arraycolsep=0.5pt
      \begin{array}{ll}
        \rho,&\xi,\\\rho',&\xi'
      \end{array}.
                            \left\{
                            \begin{array}{l}

                            \left[\bigvee_{i=0}^{n-1}\left(
                            \arraycolsep=0.5pt
                            \begin{array}{ll}
                              \rho\doteq p^{\nat}_{i}\land&\xi\doteq \natPart{x_{i}}\land\\
                              \rho'\doteq p^{\nat}_{i+1}\land&\xi'\doteq \natPart{x_{i+1}}\\
                            \end{array}
                              \right)\right]\\\implies\fraPart{\psi}\left(\rho,\xi,\rho',\xi'\right)
                            \end{array}

      \right\}.
    \end{array}
  \end{equation*}
\end{proof}

Theorem \ref{theo:bounded-construct} can now be proven.
\begin{proof}[Proof of Theorem \ref{theo:bounded-construct}]
  The algorithm is exactly the same than the algorithm of
  Proposition \ref{proposition:bounded-construct}.  It suffices to prove that
  the algorithm of Proposition \ref{proposition:bounded-construct} can be
  applied to $\mathcal{A}'$, that is, that $\mathcal{A}'\in\BAF$.
  
  Let $L=\toInfWord{\mathcal{A}}$ and $R=\wordToReal{L}$. Since $\mathcal{A}$ is fractional,
  then $L$ is also fractional, hence $L=\toInfWord R$.  By
  Lemma \ref{lem:ar-01-correct}, $L$ is also recognized by $\AR$ as in
  Definition \ref{def:AR01}. By Lemma \ref{lem:AR-in-BAF}, $\AR\in\BAF$ and
  by Lemma \ref{lem:quotient-BAF}, its minimal quotient $\mathcal{A}''$ belongs
  to $\BAF$. Since $\mathcal{A}$ and $\AR$ recognizes the same language,
  $\mathcal{A}''$ is also the minimal quotient of $\mathcal{A}$, therefore
  $\mathcal{A}''=\mathcal{A}'$ and $\mathcal{A}'\in\BAF$.
\end{proof}

\section{Conclusion}
In this paper, we proved that it is decidable in linear time whether a
weak Büchi Real Number Automaton $\mathcal{A}$ reading a set of real number
$R$ recognizes a finite union of intervals. It is proved that a
quasi-linear sized existential-universal formula defining $R$ exists.
And that a quasi-quadratic existential formula defining $R$ also
exists. 

The theorems of this paper lead us to consider two natural
generalization. We intend to adapt the algorithm of this paper to
similar problems for automata reading vectors of reals instead of
automata reading reals. We also would like to solve a similar problem,
deciding whether an RNA accepts a
$\fo{\mathbb R,\mathbb Z;+,<}$-definable set of reals. Solving this
problem would also solve the problem of deciding whether an automaton
reading natural number, beginning by the most-significant digit,
recognizes an ultimately-periodic set. Similar problems has already
been studied, see e.g
\cite{DBLP:journals/corr/cs-CC-0309052,DBLP:journals/fuin/LacroixRRV12}
and seems to be difficult.

The author thanks Bernard Boigelot, for a discussion about the
algorithm of Theorem \ref{theo:bounded-construct}, which led to a
decrease of the computation time.

\bibliographystyle{plain}
\bibliography{fo}

\begin{thebibliography}{10}

\bibitem{DBLP:journals/corr/cs-CC-0309052}
Boris Alexeev.
\newblock Minimal dfas for testing divisibility.
\newblock {\em CoRR}, cs.CC/0309052, 2003.

\bibitem{real}
Bernard Boigelot, Julien Brusten, and V{\'e}ronique Bruy{\`e}re.
\newblock On the sets of real numbers recognized by finite automata in multiple
  bases.
\newblock {\em Logical Methods in Computer Science}, 6(1), 2010.

\bibitem{weak-R-+-vector}
Bernard Boigelot, Julien Brusten, and J{\'e}r{\^o}me Leroux.
\newblock {A Generalization of Semenov's Theorem to Automata over Real
  Numbers}.
\newblock In Renate~A. Schmidt, editor, {\em {Automated Deduction, 22nd
  International Conference, CADE 2009}}, volume 5663 of {\em Lecture Notes in
  Computer Science}, pages 469--484, Montr{\'e}al, Canada, August 2009.
  {Springer Berlin}.

\bibitem{weak-buchi-r-+}
Bernard Boigelot, S{\'e}bastien Jodogne, and Pierre Wolper.
\newblock An effective decision procedure for linear arithmetic over the
  integers and reals.
\newblock {\em ACM Trans. Comput. Logic}, 6(3):614--633, July 2005.

\bibitem{bruyere}
Véronique Bruyère, Georges Hansel, Christian Michaux, and Roger Villemaire.
\newblock Logic and p-recognizable sets of integers.
\newblock {\em Bull. Belg. Math. Soc}, 1:191--238, 1994.

\bibitem{Cobham}
Alan Cobham.
\newblock On the base-dependence of sets of numbers recognizable by finite
  automata.
\newblock {\em Mathematical systems theory}, 3(2):186--192, 1969.

\bibitem{elimFOr}
Jeanne Ferrante and Charles Rackoff.
\newblock A decision procedure for the first order theory of real addition with
  order.
\newblock {\em {SIAM} J. Comput.}, 4(1):69--76, 1975.

\bibitem{number-theory-hardy}
Godfrey~Harold Hardy and Edward~Maitland Wright.
\newblock An introduction to the theory of numbers, 1960.
\newblock Autres tirages avec corrections : 1962, 1965, 1968, 1971, 1975.

\bibitem{Honkala86}
Juha Honkala.
\newblock A decision method for the recognizability of sets defined by number
  systems.
\newblock {\em ITA}, 20(4):395--403, 1986.

\bibitem{DBLP:journals/fuin/LacroixRRV12}
Anne Lacroix, Narad Rampersad, Michel Rigo, and {\'{E}}lise Vandomme.
\newblock Syntactic complexity of ultimately periodic sets of integers and
  application to a decision procedure.
\newblock {\em Fundam. Inform.}, 116(1-4):175--187, 2012.

\bibitem{Leroux}
J{\'e}r{\^o}me Leroux.
\newblock Least significant digit first {P}resburger automata.
\newblock {\em CoRR}, abs/cs/0612037, 2006.

\bibitem{minimal-buchi}
Christof Löding.
\newblock Efficient minimization of deterministic weak omega automata, 2001.

\bibitem{Sakarovitch}
Victor Marsault and Jacques Sakarovitch.
\newblock Ultimate periodicity of b-recognisable sets: A quasilinear procedure.
\newblock In Marie-Pierre B{\'e}al and Olivier Carton, editors, {\em
  Developments in Language Theory}, volume 7907 of {\em Lecture Notes in
  Computer Science}, pages 362--373. Springer, 2013.

\bibitem{muchnik}
Andrei~A. Muchnik.
\newblock The definable criterion for definability in {P}resburger arithmetic
  and its applications.
\newblock {\em Theor. Comput. Sci.}, 290(3):1433--1444, 2003.

\bibitem{semenov-theorem}
A.~L. Semenov.
\newblock The presburger nature of predicates that are regular in two number
  systems.
\newblock {\em Siberian Math. J.}, 18:289,299, 1977.

\bibitem{Tarjan}
Robert Tarjan.
\newblock Depth-first search and linear graph algorithms.
\newblock {\em SIAM Journal on Computing}, 1(2):146--160, 1972.

\bibitem{elimination}
Volker Weispfenning.
\newblock Mixed real-integer linear quantifier elimination, 1999.

\bibitem{LinArCon}
Pierre Wolper and Bernard Boigelot.
\newblock On the construction of automata from linear arithmetic constraints.
\newblock In Susanne Graf and Michael Schwartzbach, editors, {\em Tools and
  Algorithms for the Construction and Analysis of Systems}, volume 1785 of {\em
  Lecture Notes in Computer Science}, pages 1--19. Springer Berlin Heidelberg,
  2000.

\end{thebibliography}
\printindex
\end{document}